\documentclass[11pt,a4paper]{article}
\usepackage{latexsym}

\newcommand{\beq}{\begin{equation}}
\newcommand{\eeq}{\end{equation}}
\newcommand{\w}{\omega}
\newcommand{\vecr}{\mathbf{r}}

\renewcommand{\baselinestretch}{1.3} 
\renewcommand{\theequation}{\thesection.\arabic{equation}}

\begin{document}
\pagenumbering{roman} \setcounter{page}{0} 
\title{Four-dimensional polymer collapse II:\\
Pseudo-First-Order Transition in Interacting Self-avoiding Walks}
\author{T. Prellberg\dag \, and A. L. Owczarek\ddag\thanks{{\tt {\rm email:}
tprell@physics.syr.edu, aleks@ms.unimelb.edu.au}} \\
	\dag Department of Physics,\\
	Syracuse University, Syracuse, NY 13244, USA\\
         \ddag Department of Mathematics and Statistics,\\
         The University of Melbourne,\\
         Parkville, Victoria 3052, Australia.}
\date{\today}

\maketitle 
 
\begin{abstract} 

In earlier work we provided the first evidence that the collapse, or
coil-globule, transition of an isolated polymer in solution can be
seen in a four-dimensional model. Here we investigate, via Monte Carlo
simulations, the canonical lattice model of polymer collapse, namely
interacting self-avoiding walks, to show that it not only has a
distinct collapse transition at finite temperature but that for any
finite polymer length this collapse has many characteristics of a
rounded first-order phase transition. However, we also show that there
exists a `$\theta$-point' where the polymer behaves in a simple
Gaussian manner (which is a critical state), to which these
finite-size transition temperatures approach as the polymer length is
increased. The resolution of these seemingly incompatible conclusions
involves the argument that the first-order-like rounded transition is
scaled away in the thermodynamic limit to leave a mean-field
second-order transition. Essentially this happens because the
finite-size
\emph{shift} of the transition is asymptotically much larger than the
\emph{width} of the pseudo-transition and the latent heat decays to
zero (algebraically) with polymer length. This scenario can be
inferred from the application of the theory of Lifshitz, Grosberg and
Khokhlov (based upon the framework of Lifshitz) to four dimensions:
the conclusions of which were written down some time ago by
Khokhlov. In fact it is precisely above the upper critical dimension,
which is $3$ for this problem, that the theory of Lifshitz may be
quantitatively applicable to polymer collapse.

\vspace{1cm} 

\noindent{\bf Short Title:} 4d ISAW


\end{abstract} 

\vfill

\newpage

\pagenumbering{arabic}
\section{Introduction} 
\setcounter{page}{1}

The collapse or coil-globule transition of an isolated polymer in
solution has been studied by a variety of different theoretical
approaches over the past 50 years ranging from phenomenological
arguments, field theoretic renormalisation group approaches, continuum
path integrals and the analysis of discrete lattice walks
\cite{cloizeaux1990a-a}. Application and testing of these theories has
mainly been confined to the ``physical'' dimensions of two and
three. However, the phase transition of polymer collapse has been long
believed to have an upper critical dimension of three ($d_u=3$) and so
the differences between the predictions of many of the theories in
that dimension lie in subtle logarithmic factors that are difficult to
ascertain numerically
\cite{grassberger1995a-a}. In two dimensions the field theoretic
(excluding conformal field theories) and continuum models have not
given exact answers and cannot be compared to the conjectured
\cite{duplantier1987a-a} and
numerically resolved \cite{prellberg1994a-:a} values of universal
quantities. Until recently
\cite{owczarek1998a-:a} four dimensions has largely been ignored but
here we argue that not only may four-dimensional studies be important to
delineate which theoretical descriptions are valid, but that the
collapse transition in four dimensions has some intriguing features of
general interest in the field of phase transition and critical
phenomena in statistical mechanics.

An isolated polymer in solution is usually considered to be in one of
three states depending on the strength of the inter-monomer
interactions which are mediated by the solvent molecules and can be
controlled via the temperature $T$. At high temperatures and in so
called ``good solvents'' a polymer chain is expected to be in a
swollen phase (swollen coil) relative to a reference Gaussian state so
that the average size of the polymer scales with chain length
algebraically faster than it would if it were behaving as a random
walk. At low temperatures or in poor solvents the polymer is expected
to be in a collapsed globular form with a macroscopic density inside
the polymer.  This implies an average size that scales slower than a
random walk. Between these two states there is expected to be a
second-order phase transition (sharp in the infinite chain length
limit).

The standard description of the collapse transition is a tricritical
point related to the $n \rightarrow 0$ limit of the
$(\phi^2)^2$--$(\phi^2)^3$ O($n$) field theory
\cite{gennes1975a-a,stephen1975a-a,duplantier1982a-a}. One might then
expect that above the upper critical dimension ($d_u=3$) some type of
self-consistent mean-field theory based upon a suitable tricritical
Landau-Ginzberg Hamiltonian \cite{lawrie1984a-a} would give a full
description of the transition, and hence conclude that in all
dimensions $d> 3$ there is a collapse transition from a swollen state
to the globular state with classical tricritical behaviour. There have
been various other mean-field type approaches to this problem though
their conclusions in three dimensions are similar
\cite{gennes1975a-a,lifshitz1968a-a} to each other.

The application of the mean-field theory of a tricritical point to
polymer collapse predicts that at the transition point the polymer
behaves as a random walk ($\nu=1/2$), and this point has been known as
the $\theta$-point (the $\theta$-point was originally defined as the
point where the second virial coefficient of a dilute solution of
polymers is zero, though it is expected that these definitions are
equivalent). Thermodynamically, one expects a weak transition with a
jump in the specific heat $\alpha=0$ (note that the thermodynamic
polymer exponent $\alpha$ is related to the shift exponent
$\psi=2-\alpha$ in tricritical theory \cite{lawrie1984a-a}, itself not
to be confused with the polymer theory finite-size scaling shift
exponent). For finite polymer length $N$ there is no sharp transition
for an isolated polymer (unless one examines a macroscopic number of
such polymers) and so this mean-field transition is rounded. In three
dimensions the application of various self-consistent mean-field like
approaches leads to the prediction that the second-order transition is
rounded and shifted on the same scale of $N^{-1/2}$, that is, the
crossover exponent $\phi$ is $1/2$, though strictly the power laws
involved are modified via renormalisation group arguments
\cite{duplantier1986b-a,duplantier1987d-a} by confluent
logarithms. (In particular note that it is predicted that in three
dimensions the specific heat should be divergent logarithmically.) In
four and higher dimensions no confluent logarithms should be present
and one may expect pure mean-field behaviour with a crossover exponent
of $1/2$ ($\phi_t$ is the relevant tricritical exponent here
\cite{lawrie1984a-a}). 

On the other hand, Domb \cite{domb1974a-a} suggested some time ago
that polymer collapse may be a first-order transition in three
dimensions and the analysis by de Gennes \cite{gennes1975a-a} of the
three-dimensional case of a suitably extended (Flory-type)
self-consistent mean-field approach predicted in some parameter
regions a first-order transition: this was superseded by his
renormalisation group approach
\cite{gennes1975a-a,gennes1978a-a}. In contrast some time ago there was the
conjecture that the collapse transition disappears altogether above
three dimensions, at least at finite temperature
\cite{moore1976a-a}. For $d>3$ Sokal \cite{sokal1994a-a} has
also pointed out that the field theoretic/Edwards model approaches
have difficulties: in fact, if one analyses the Edwards model one
finds the crossover exponent is given by $\phi=2-d/2$, which for $d=4$
gives $\phi=0$! In passing we note here that the same analysis
predicts the shift of the $\theta$-point, defined say via the
universal ratio of the radius of gyration to the end-to-end distance
equalling its Gaussian value, should scale as $N^{-(d/2 -1)}$ which
has (polymer) shift exponent $1$ in dimension $d=4$. This difference
between the shift and the crossover exponent implies that strict
crossover scaling has broken down. Of course, the theoretical fact that
the swollen phase should also be Gaussian for $d>4$ does raise the
suspicion that the analysis of the Edwards model for polymer collapse
may be subtle for $d>3$.

As a first attempt to explore the issues raised above we recently
\cite{owczarek1998a-:a} considered the problem of interacting
self-avoiding trails on the four-dimensional hyper-cubic lattice with
a special set of Boltzmann weights as generated by a kinetic growth
algorithm.  Interacting self-avoiding trails are a candidate lattice
model for polymer collapse. They are defined to be lattice paths such
that each bond of the lattice may either be unoccupied or occupied by
a single bond of the path, though sites can be multiply
occupied. Attractive interactions are associated with those multiply
visited sites so that the strength of this interaction drives a
collapse. The study had several virtues. Firstly, since trails are
allowed to intersect, but still possess excluded volume, we were able
to grow configurations without the normal attrition that appears in
growing self-avoiding walks through the ``trapping'' of the growth in
dense sections of the polymer. Secondly, in lower dimensions it had
been seen that whenever a kinetic growth algorithm was un-hindered by
trapping it mapped precisely onto the collapse transition point of the
model
\cite{bradley1989a-a,bradley1990a-a,prellberg1994a-:a,prellberg1995b-:a}. In
\cite{owczarek1998a-:a} we indeed found a set of Boltzmann weights where
the model appears to have the Gaussian characteristics of a
$\theta$-point. The two drawbacks of this approach are, firstly, that
the simulations cannot effectively be extended away from the special
temperature and, secondly, that the canonical lattice model of polymer
collapse is rather self-avoiding walks (SAW) interacting via
nearest-neighbour site (monomer) attraction. This canonical model is
known as interacting self-avoiding walks, hence ISAW. A new algorithm,
known as PERM, for the Monte Carlo simulation of ISAW (among other
things) has been recently developed by Grassberger and collaborators
\cite{grassberger1997a-a,frauenkron1998a-a,caracciolo1999a-a}. This is
essentially a kinetic growth 
strategy that adds clever enhancements to simultaneously allow a wide
range of temperatures to be accessed and for the attrition of samples
through trapping to be lessened (see section \ref{perm} for a fuller
description). In this paper we have simulated ISAW on the
four-dimensional hyper-cubic lattice using the PERM algorithm over a
wide range of temperatures.

Now, to begin, our results suggest that there is indeed a collapse
transition in four dimensions at a finite temperature. However, the
character of that transition is particularly intriguing! We find a
distinct double peak distribution for the internal energy far below a
point which we clearly identify as a candidate $\theta$-point. This
double peak distribution becomes \emph{more} pronounced as the chain
length is increased.  This would seem to suggest a first-order
transition. If this was the case there would be a delta function peak
forming in the specific heat but we find that while a peak is indeed
forming it seems not to be growing linearly with the size of the
polymer. More importantly, the location of a distinct $\theta$-point
is incompatible with a first-order transition if there is only one
collapse! However there is a theoretical framework (whose conclusions
are suitably extended here) that is consistent with the evidence we
present. This framework was explained in a paper by Khokhlov
\cite{khokhlov1981a-a} who applied the mean-field approach of
Lifshitz, Grosberg and Khokhlov (LGK)
\cite{lifshitz1968a-a,lifshitz1976a-a,lifshitz1978a-a} to arbitrary
dimensions\footnote{We warn the reader that the abstract (and parts of
the conclusions) of the paper
\cite{khokhlov1981a-a} may be misleading as it reads that ``for $d>3$
the coil-globule transition is of first-order''. Moreover the
application of this mean-field theory to $d<3$ is inappropriate and
its conclusions have now been superseded.}.  The LGK theory is based
on a phenomenological free energy in which the competition between a
bulk free energy of a dense globule and its surface tension drive the
transition. Until recently \cite{owczarek1993c-:a} the consequences of
this surface free energy were largely ignored in the polymer
literature. Its effect on the scaling form of the finite-size
partition function was argued and confirmed
\cite{owczarek1993c-:a,owczarek1993d-:a,grassberger1995a-a,nidras1996a-a}.
We shall refer to the LGK theory as applied to dimensions four and
above as KLG to distinguish it from the original three-dimensional
work of Lifshitz, Grosberg and Khokhlov
\cite{lifshitz1976a-a,lifshitz1978a-a}. 

Hence the major conclusions of our work are that the finite-size
character of the coil-globule transition in four dimensions is
first-order despite the thermodynamic limit being probably adequately
described by mean-field tricritical behaviour. The only alternative
conclusion from our data is that the transition is truly first-order and
our finding of a $\theta$-state is fortuitous. The whole theory of
crossover scaling for this transition needs to be reworked. This
curious state of affairs where a second-order transition looks
distinctly first-order may be of interest in other physical situations
where mean-field theory is used to describe thermodynamics.

The layout of the paper is as follows. In the next section we define
the model we consider and review the generally expected behaviour of
the quantities we have calculated in four dimensions. Then in
section 3 we explain the results of KLG theory as applied to four
dimensions. In section 4 we explain our Monte Carlo approach, PERM, and 
finally in section 5 we carefully describe the numerical results of
our simulations and how well they conform to the theory of KLG.

\section{The ISAW model and a review of basic scaling results}

	The interacting self-avoiding walk model is the canonical
lattice model of the coil-globule transition and has been long studied 
in two and three dimensions. Here we shall consider the
four-dimensional hyper-cubic lattice  (coordination number 8). The
monomers are imagined to be sitting on the sites of the lattice and a
self-avoiding path of such sites form the polymer. The self-avoidance
means that no two monomers can sit at the same site of the lattice. 

        The partition function of the self-interacting self-avoiding
walk model (ISAW) is given by
\begin{equation}
Z_N(\w) = \sum_{\varphi\in\Omega_N} \w^{m(\varphi)} \; ,
\end{equation}
where the sum is over the set of all self-avoiding walks $\Omega_N$ of
length $N$ steps ($N+1$ monomers) with one end at some fixed origin
and $m(\varphi)$ is the number of non-consecutive nearest-neighbour
monomers for a given walk $\varphi$. The Boltzmann weight 
$\w=e^{\beta\epsilon}$ is associated
with a nearest-neighbour contact of energy $-\epsilon$ so that $\omega
> 1.0$ for attractive interactions. We define a reduced finite-size
free energy per step $\kappa_N(\omega)$ as
\begin{equation}
\kappa_N(\omega)={1\over N}\log Z_N(\omega).
\end{equation}
The usual free energy is related to this by $-\beta F_N\equiv N\kappa_N(\omega)$.

The average of any quantity $Q$ over the
ensemble set of allowed paths $\Omega_N$ of length $N$ is given
generically by
\begin{equation}
\langle Q \rangle_N (\omega)= \frac{\sum_{\varphi\in\Omega_N}
Q(\varphi) \w^{m(\varphi)}}{\sum_{\varphi\in\Omega_N} \w^{m(\varphi)}}\; .
\end{equation}
We define a normalised finite-size internal energy per step by
\begin{equation}
U_N(\omega) = \frac{\langle m \rangle}{N}\, ,
\end{equation}
and a normalised finite-size specific heat per step by
\begin{equation}
C_N(\omega) = \frac{\langle m^2 \rangle - \langle m \rangle^2}{N}\; .
\end{equation}
These quantities are related in the usual way to the reduced free energy via
$U_N=\partial\kappa_N/\partial\log\omega$ and $C_N=\partial U_N/\partial\log\omega$.

The thermodynamic limit in this problem is given by the limit 
$N\rightarrow\infty$ so that the thermodynamic free energy per 
step $f_{\infty}(\w)$ is given by
\begin{equation}
-\beta f_{\infty}(\w)= \kappa_\infty(\omega)=\lim_{N \rightarrow \infty}\kappa_N(\omega) \; .
\end{equation}
This quantity determines the partition function asymptotics, i.e. $Z_N(\omega)$ grows to
leading order exponentially as $\mu(\omega)^N$ with $\mu(\omega)=e^{\kappa_\infty(\omega)}$.

In our simulations we calculated two measures of the polymer's average
size.  Firstly, specifying a walk by the sequence of position vectors
${{\bf r}_0, {\bf r}_1, ..., {\bf r}_N}$ the average mean-square
end-to-end distance is
\begin{equation}
\langle R^2_e\rangle_N = \langle ({\bf r}_N- {\bf r}_0) \cdot ({\bf r}_N- {\bf r}_0) \rangle\; .
\end{equation} 
We shall use the symbol $R^2_{e,N}$ to be equivalent to 
\begin{equation}
R^2_{e,N} (\omega)\equiv \langle R^2_e\rangle_N .
\end{equation} 
The mean-square distance of a monomer from the endpoint $\vecr_0$ is
given by 
\begin{equation}
\langle R_{m}^2 \rangle_N = \frac{1}{N+1}\sum_{i=0}^{N}\langle (\vecr_i - \vecr_0)\cdot(\vecr_i - \vecr_0)
\rangle\; .
\end{equation}
Again we define
\begin{equation}
R^2_{m,N}(\omega) \equiv \langle R^2_m\rangle_N.
\end{equation}
We also define the ratio 
\begin{equation}
B_N(\omega) = \frac{R^2_{m,N}}{R^2_{e,N}},
\end{equation}
which should have a universal limit in each
critical phase of the model.

Now let us \emph{assume} for a moment that there is a single collapse
transition at some value of $\omega$ and let us explore the
(four-dimensional) behaviour we might expect from the above defined
quantities in each of the phases.  As discussed above, the basic
physics of the coil-globule (collapse) transition can be understood by
the consideration of the average size of the polymer, $R_N$, either $R_{e,N}$
or $R_{m,N}$, as a function of length $N$ in each of the phases, so let
us consider this first. Generally one always expects that
\begin{equation}
R^2_{N} \sim a(\omega)\: N^{2\nu}\quad\mbox{as}\quad N\rightarrow\infty
\end{equation}
for any fixed temperature.  In four dimensions at infinite temperature,
$\omega=1$, it has been predicted \cite{gennes1979a-a} that
\begin{equation}
\label{swollen-size}
R^2_{N} \sim a^{+}\: N \left(\log(N)\right)^{1/4} .
\end{equation}
If there does exist a collapse transition then one would expect that
this scaling extends (with a constant $a^{+}$ that depends on
temperature) down to the transition point. In the collapsed phase the
polymer is expected to assume a dense configuration on average and
hence the globular value of the radius-of-gyration exponent is
$\nu_g=1/d=1/4$ \cite{gennes1975a-a} with
\begin{equation}
\label{collapse-size}
R^2_{N} \sim a^{-}(\omega)\: N^{1/2}.
\end{equation}
Finally at some finite
transition temperature $1.0 < \omega_t < \infty$ a Gaussian scaling of
the radius of gyration should occur, that is
\begin{equation}
\label{theta-size}
R^2_{N} \sim a^\theta\: N,
\end{equation}
so that $\nu_t=1/2$. This Gaussian scaling is often used
(theoretically at least) to define the $\theta$-point
$\omega=\omega_\theta$ of an isolated polymer so that
$\omega_t=\omega_\theta$. The universal ratio $B_N$ is expected to
converge to the value $B_\infty= 1/2$ both in the swollen phase and at
$\omega_\theta$. However, one would expect slow logarithmic
corrections for $\omega < \omega_\theta$ and algebraic corrections at
$\omega_\theta$. For $\omega > \omega_\theta$ the phase is no longer
expected to be critical and so $B_\infty$ is no longer universal and
may be a non-constant function of $\omega$.

One can also consider the scaling of the partition function in each of
the regimes, given that there is a transition. For high temperatures
$1.0 < \omega < \omega_\theta$ one expects the infinite temperature
behaviour, which is
\cite{gennes1979a-a}
\begin{equation}
\label{swollen-partition-function}
Z_N \sim b^{+}(\omega)\: \mu(\omega)^N \:\left(\log N\right)^{1/4},
\end{equation}
while at low temperatures \cite{owczarek1993c-:a} one expects
asymptotics of the form
\begin{equation}
\label{collapse-partition-function}
Z_N \sim b^{-}(\omega)\: \mu(\omega)^N\: \mu_s(\omega)^{N^{3/4}}\, N^g
\end{equation}
where $\mu_s$ is related to the surface free energy of the polymer
globule and the exponent $g$ need not be universal (we only write it
for completeness of the asymptotic form).  For $\omega=\omega_\theta$
one expects
\begin{equation}
\label{theta-partition-function}
Z_N \sim b^{\theta}\: \mu(\omega_\theta)^N
\end{equation}
as a reflection of Gaussian behaviour.

In the thermodynamic limit the thermodynamic functions
$f_{\infty}(\omega)$, $U_{\infty}(\omega)$ and $C_{\infty}(\omega)$
are all expected to be analytic functions of $\omega$ except at
$\omega_\theta$. By using the correspondence to the tricritical model
\cite{gennes1975a-a} the mean field theory would imply that the
specific heat had a jump discontinuity at $\omega_\theta$ since
$\alpha=2-\psi=0$.  Of course, for finite $N$ there is no sharp
transition for an isolated polymer (unless one examines a macroscopic
number of such polymers). Around the collapse temperature in
\emph{three} dimensions the finite-size corrections are expected to
take on a cross-over scaling form \cite{gennes1975a-a} so that
\begin{equation}
\label{scaling-form}
R^2_N \sim a^\theta\; N \:\mathcal{R}((T-T_\theta)N^{\phi})
\end{equation}
with $\phi=1/2$ but this form implies that the transition is rounded
and shifted on the \emph{same} scale of $N^{-\phi}$ and so its
applicability to four dimensions needs careful thought.

\section{The theory of Khokhlov, Lifshitz, and Grosberg (KLG)}

We now review the results of Khokhlov \cite{khokhlov1981a-a} paying
special attention to the predicted behaviour of the quantities we have
calculated.  The theory originally proposed by Lifshitz for the
general mean-field description of the globular state, extended by
Lifshitz, Grosberg and Khokhlov to fully describe the transition, and
finally applied to four dimensions by Khokhlov \cite{khokhlov1981a-a}
is based on several phenomenological mean-field assumptions. Firstly,
there exists a state where the excluded volume property of long chain
molecules is exactly cancelled by the attractive interactions between
parts of the polymer as mediated by the solvent. This is the
$\theta$-state. Secondly, when the attraction becomes even stronger
there eventuates a globular state where the polymer behaves as a
liquid drop. The results of the theory are based on a phenomenological
free energy of that globular state relative to the free energy of the
pure Gaussian state of the $\theta$-point at $T_\theta$. Hence the
condition applied to find the finite-size position of the transition
is to equate the relative free energy to zero. The relative free
energy is given as
\begin{equation}
F_N = F_{bulk} + F_{surface}
\end{equation}
where the $F_{bulk}$ and $F_{surface} $ are given in terms of the
second and third virial coefficients, the length of the chains, and
the linear size of the polymer found from the globular density. In
particular both the bulk and surface free energies are proportional to
the square of the second virial coefficient.
It is assumed that on approaching the $\theta$-point the second virial
coefficient goes to zero linearly with temperature while the third
virial coefficient remains non-zero. Note that this implies a quadratic
dependence of the bulk free energy on the distance to the
$\theta$-point. Since the free energy has exponent $2-\alpha$ this
implies an exponent $\alpha=0$ (assuming that this part of the free
energy is singular). Therefore a second-order phase transition occurs
in the thermodynamic limit.

It is further assumed that the density in the globule is proportional
to the second virial coefficient and hence also goes to zero linearly
with temperature on approaching the $\theta$-point ($\beta=1$). By
applying the condition $F_N=0$, Khokhlov \cite{khokhlov1981a-a} finds
a finite-size transition temperature $T_{c,N}$ that approaches the
$\theta$-temperature from below as $N^{-1/3}$.  Applying this result
to the ISAW model gives
\begin{equation}
\label{shift}
\omega_{c,N} -\omega_\theta  \sim \frac{s}{N^{1/3}}
\end{equation}
for some constant $s$. That is, the polymer collapse \emph{shift}
exponent is $1/3$.

Now Khokhlov \cite{khokhlov1981a-a} found that the free energy (for $T
< T_\theta$) can be rewritten in terms of $T_{c,N}$ as
\begin{equation}
\label{KLG-free-energy}
F_N \sim  - N T \left( T_\theta - T \right)^2 \left( 1- \left|\frac{T_\theta - T_{c,N}}{T_\theta - T}\right|^{3/4} \right).
\end{equation}
Actually KLG's unwritten assumption is that $F_N$ is the minimum of
zero and the right-hand side of (\ref{KLG-free-energy}) so it is zero
for $T_\theta<T<T_{c,N}$.  Khokhlov \cite{khokhlov1981a-a} deduces
that the width of the transition $\Delta T$ at finite $N$ can be found
from this and scales as $N^{-2/3}$.
Hence the ISAW model should have a transition width $\Delta \omega$
that scales as
\begin{equation}
\label{width}
\Delta \omega \sim \frac{w}{N^{2/3}}
\end{equation}
for some constant $w$. That is, the polymer collapse \emph{crossover}
exponent is $2/3$. Hence note that the size of the crossover region is
asymptotically small relative to the shift of the transition, and that
crossover forms such as (\ref{scaling-form}) may not be useful.

	Even though there can be no sharp transition for a single
polymer of finite length $N$ the theory can describe the nature of the
rounded transition by considering the difference between the density
of the globular state $\rho_g$ at $T_{c,N}$ and that of the coil state
$\rho_c$ at the
same temperature. This is 
\begin{equation}
\frac{\rho_g(T_{c,N}) - \rho_c(T_{c,N})}{\rho_c(T_{c,N})} \propto N^{2/3},
\end{equation}
which diverges as $N$ becomes large. Hence Khokhlov
\cite{khokhlov1981a-a} concluded that `the 
coil-globule transition is first-order', though we now interpret this
to mean that the finite-size corrections to the thermodynamic
second-order transition are first-order like\footnote{We point the
terminology of Khokhlov was presumably that explained in Section I.C.2
of \cite{lifshitz1978a-a} but may be misleading to the modern
reader.}. However, both $\rho_g(T_{c,N})$ and $\rho_c(T_{c,N})$ tend
to zero as $N \rightarrow
\infty $ and it is simply that $\rho_g(T_{c,N})$ tends to zero
asymptotically slower than $\rho_c(T_{c,N})$ that makes the relative
difference quoted above diverge. Noting that $\rho=N/R^4$ and $R_c
\sim N^{1/2}$, the  above equation can be used to deduce the scaling of
$R_N$ at $T_{c,N}$ as 
\begin{equation}
\label{size-at-collapse}
R_N (T_{c,N}) \sim a^c N^{1/3}.
\end{equation}
Hence we define an effective radius-of-gyration exponent $\nu_c=1/3$
for the scaling of the size of the polymer when following the
finite-size transition temperatures. Note that this exponent value
obeys $\nu_\theta=1/2 >\nu_c > \nu_g=1/4$.

	Following the the work of Lifshitz, Grosberg and Khokhlov
\cite{lifshitz1978a-a} one can also calculate the change in the
internal energy over the crossover width of the transition $\Delta T$
as the latent heat (or ``heat of the transition'') by using expression
(\ref{KLG-free-energy}):
\begin{equation}
\label{latent-heat}
\Delta U \sim  \frac{u^c}{N^{1/3}}.
\end{equation}
The corresponding height of the peak in the specific heat is
\begin{equation}
\label{specific-heat-peak}
C_{N}(T_{c,N}) \sim h^c\; N^{1/3}.
\end{equation}

So to summarise the picture so far, the theory predicts a
thermodynamic second-order transition at a $\theta$-point with a jump
in the specific heat. For finite polymer length this transition is
shifted below the $\theta$-point by a temperature of the order of
$O(N^{-1/3})$ with the width of the transition of the order of
$O(N^{-2/3})$. Over this width there is a rapid change in the internal
energy that scales as $O(N^{-1/3})$: the important point here of
course is that this tends to zero for infinite length so the effect of
the peak in the specific heat is scaled away for $N$ large, leaving a
finite jump in the thermodynamic limit. To understand this further let
us consider the distribution of internal energy as a function of
temperature and length. For any temperature above the $\theta$-point
and well below $T_{c,N}$ one expects the distribution of internal
energy to look like a single peaked distribution centred close to the
thermodynamic limit value: a Gaussian distribution is expected around
the peak with variance $O(N^{-1/2})$. In fact, this picture should be
valid for all temperatures outside the range $[T_{c,N} -
O(N^{-2/3}),T_{c,N} + O(N^{-2/3})]$. When we enter this region we
expect to see a double peaked distribution as in a first-order
transition region. For any temperature in this region there should be
two peaks in the internal energy distribution separated by a gap
$\delta U$ of the order of $\delta U\approx \Delta U \propto
O(N^{-1/3})$. Each peak is of Gaussian type with individual variances
again of the order of $O(N^{-1/2})$. Hence as $N$ increases the peaks
will become more and more distinct and relatively sharper but the peak
positions will be getting closer together. We refer to this scenario
as a \emph{pseudo}-first-order transition or, more correctly, as
first-order-like finite-size corrections to a second-order phase
transition. If there were a real first-order transition then the
distance between the peaks should converge to a non-zero constant.

\section{Perm} 
\label{perm}

We have simulated ISAW using the Pruned-Enriched Rosenbluth Method
(PERM), a recently proposed generalisation of a simple kinetic growth
algorithm \cite{grassberger1997a-a,frauenkron1998a-a}.  PERM builds
upon the Rosenbluth-Rosenbluth method \cite{rosenbluth1955a-a}, in
which walks are generated by simply growing an existing walk
kinetically, i.e.  by choosing the next step with equal probability
from all possible accessible lattice sites. Eventually, a walk
generated thus gets trapped in a configuration in which it cannot be
continued, leading generally to an exponential ``attrition''.
Moreover, in order to simulate ISAW at a particular temperature one
needs to re-weight the kinetically grown samples in such a way that
the generated sample is usually dominated by a few configurations
which carry large weight after the re-weighting.

In order to overcome these obstacles, PERM uses a combination of
enrichment and pruning strategies to generate walks whose weights are
largely distributed around the expected peak of the distribution. On
the one hand, if the weight of a configuration becomes too small, the
configuration gets pruned probabilistically and the weight adjusted
correspondingly. Alternatively, if the weight of a configuration
becomes too large, copies of the walk are made and the respective
weights reduced accordingly. While this does not eliminate trapping,
it is generally sufficient to compensate for it: trapping occurs when
the end of the walk is in an area of high density, which in turn
increases the likelihood of enrichment. The algorithm can be
implemented in a self-tuning way by choosing dynamically adjusted
upper and lower threshold values to control pruning and enrichment
rates.

It is plausible that the algorithm works best at temperatures in which
the thermal distribution is close to the distribution of walks
generated by kinetic growth. In sufficiently large dimensions, this
temperature should be quite close to the $\theta$-temperature, so that
the algorithm is expected to be well suited to the study of polymer
collapse. As mentioned in the introduction, there are lattice models
of interacting polymers for which there is an exact mapping of the
corresponding kinetic growth models to their respective $\theta$-points,
in which case a PERM simulation at the $\theta$-temperature reduces to
simple kinetic growth. In fact, it turns out that the algorithm
performs well over a whole range of temperatures covering all of the
swollen phase and the scaling region around the collapse
transition. However, we find that the performance of PERM in the
collapsed phase is far less satisfactory.

The guiding principle for any choice of implementation should be the
observation \cite{grassberger1997a-a} that the algorithm essentially
produces a random walk in chain length with reflecting boundaries at
$0$ and $N_{max}$. Considered in such a way, the algorithm performs
best if this random walk is unbiased and if the associated diffusion
coefficient is large. To eliminate bias, pruning, trapping and
enrichment rates have to compensate each other. To maximise the
diffusion coefficient, the pruning and trapping rates have to be as
small as possible. The choice of pruning and enrichment thresholds
needs to take both into account. (In contrast with the original work
on PERM, where trapping was viewed as a special case of pruning, we
find it instructive to distinguish between these two effects: trapping
is unavoidable due to the geometry of the lattice, whereas pruning is
done optionally to adjust weights.)

In our implementation, we chose upper and lower thresholds $W^u$ and
$W^l$ proportional to the current estimate of the average weight of a
walk at length $N$, $\langle Z_N\rangle/s_N$, where $s_N$ is the
number of generated samples at length $N$, and $\langle Z_N\rangle$ is
the current estimate of the partition function at length $N$. That is
to say,
\beq
W^u_N=c^u_N\langle Z_N\rangle/s_N\;,\quad W^l_N=c^l_N\langle
Z_N\rangle/s_N\; .
\eeq
In order to enforce an even sample size distribution we allow for
dynamic adjustment of $c^u_N$ and $c^l_N$. Thus, if for example at any
particular length we have an excess of pruning, the algorithm
``relents'' and increases both $c^u_N$ and $c^l_N$ in order to reduce
pruning and enhance enrichment, keeping the quotient of the thresholds
$Q=c^u_N/c^l_N$ constant. To stabilise the dynamic adjustment, we
enforce $c^u_N>c^u_{min}$ and $c^l_N<c^l_{max}$. After some initial
experimentation, we chose $c^u_{min}=2$ and $c^l_{max}=1/2$, which
leaves us with the threshold quotient $Q$ as the sole adjustable
parameter. (We also experimented with a dynamic length-dependent
$Q_N$, but the dynamic adjustment seemed to be too unstable to pursue
this avenue further.) For each run, we attempted to choose the
smallest threshold quotient $Q$ for which we could obtain an even
sample size distribution.

The disadvantage of PERM is that due to the enrichment the generated
data is not independent. All the data generated during one ``tour'',
i.e. between two successive returns of the algorithm to length $0$, is
correlated. Therefore, we keep track of the statistics of tour sizes
$t$ to get a rough idea of the quality of the data. In our
statistical evaluation we use (somewhat arbitrarily) the quotient of
$s_N$ and $\sqrt{\langle t^2\rangle}$ as a measure of an effective
independent sample size. This is correct as long as the tour sizes
don't fluctuate too strongly, and, more importantly, as long as
individual tours explore the sample space evenly. When simulating in
the collapsed phase, both of these assumptions break down, and the
sample is dominated by few huge tours. Moreover, the pruning and
enrichment rates become so large that the efficiency of the algorithm
is significantly decreased.

For further details of the algorithm and suggestions of various other
improvements such as Markovian anticipation we refer to
\cite{frauenkron1998a-a,caracciolo1999a-a}.

\section{Results} 

We simulated ISAW on a 4-dimensional hyper-cubic lattice using PERM
with $N_{max}$ set to 1024, 2048, 4096, 8192, and 16384 at values of
$\omega$ ranging from $1.0$ to $1.4$ ($N_{max}=1024$) to $1.2175$ to
$1.2225$ ($N_{max}=16384$) and the threshold quotient $Q$ ranging from
$10$ to $160$.  At each fixed $\omega$, we generated $10^7$ walks. To
illustrate the computational effort, the generation of a sample of
size $10^7$ at length $N_{max}=16384$ took about $2$ weeks CPU time on
a $600$ MHz DEC Alpha.

We computed statistics for $R_{e,N}^2$ and $R_{m,N}^2$, the partition
function $Z_N$, the internal energy $U_N$ and specific heat
$C_N$. Moreover, we generated the distribution of the number of
interactions at $N_{max}$. The distributions obtained at various
temperatures were then combined using the multiple histogram method
\cite{ferrenberg1988a-a}.

\begin{figure}[p]
\begin{center}
\setlength{\unitlength}{0.240900pt}
\ifx\plotpoint\undefined\newsavebox{\plotpoint}\fi
\sbox{\plotpoint}{\rule[-0.200pt]{0.400pt}{0.400pt}}
\begin{picture}(1500,900)(0,0)
\font\gnuplot=cmr10 at 10pt
\gnuplot
\sbox{\plotpoint}{\rule[-0.200pt]{0.400pt}{0.400pt}}
\put(161.0,123.0){\rule[-0.200pt]{4.818pt}{0.400pt}}
\put(141,123){\makebox(0,0)[r]{0.0}}
\put(1419.0,123.0){\rule[-0.200pt]{4.818pt}{0.400pt}}
\put(161.0,270.0){\rule[-0.200pt]{4.818pt}{0.400pt}}
\put(141,270){\makebox(0,0)[r]{0.5}}
\put(1419.0,270.0){\rule[-0.200pt]{4.818pt}{0.400pt}}
\put(161.0,418.0){\rule[-0.200pt]{4.818pt}{0.400pt}}
\put(141,418){\makebox(0,0)[r]{1.0}}
\put(1419.0,418.0){\rule[-0.200pt]{4.818pt}{0.400pt}}
\put(161.0,565.0){\rule[-0.200pt]{4.818pt}{0.400pt}}
\put(141,565){\makebox(0,0)[r]{1.5}}
\put(1419.0,565.0){\rule[-0.200pt]{4.818pt}{0.400pt}}
\put(161.0,713.0){\rule[-0.200pt]{4.818pt}{0.400pt}}
\put(141,713){\makebox(0,0)[r]{2.0}}
\put(1419.0,713.0){\rule[-0.200pt]{4.818pt}{0.400pt}}
\put(161.0,860.0){\rule[-0.200pt]{4.818pt}{0.400pt}}
\put(141,860){\makebox(0,0)[r]{2.5}}
\put(1419.0,860.0){\rule[-0.200pt]{4.818pt}{0.400pt}}
\put(161.0,123.0){\rule[-0.200pt]{0.400pt}{4.818pt}}
\put(161,82){\makebox(0,0){1.00}}
\put(161.0,840.0){\rule[-0.200pt]{0.400pt}{4.818pt}}
\put(321.0,123.0){\rule[-0.200pt]{0.400pt}{4.818pt}}
\put(321,82){\makebox(0,0){1.05}}
\put(321.0,840.0){\rule[-0.200pt]{0.400pt}{4.818pt}}
\put(481.0,123.0){\rule[-0.200pt]{0.400pt}{4.818pt}}
\put(481,82){\makebox(0,0){1.10}}
\put(481.0,840.0){\rule[-0.200pt]{0.400pt}{4.818pt}}
\put(640.0,123.0){\rule[-0.200pt]{0.400pt}{4.818pt}}
\put(640,82){\makebox(0,0){1.15}}
\put(640.0,840.0){\rule[-0.200pt]{0.400pt}{4.818pt}}
\put(800.0,123.0){\rule[-0.200pt]{0.400pt}{4.818pt}}
\put(800,82){\makebox(0,0){1.20}}
\put(800.0,840.0){\rule[-0.200pt]{0.400pt}{4.818pt}}
\put(960.0,123.0){\rule[-0.200pt]{0.400pt}{4.818pt}}
\put(960,82){\makebox(0,0){1.25}}
\put(960.0,840.0){\rule[-0.200pt]{0.400pt}{4.818pt}}
\put(1120.0,123.0){\rule[-0.200pt]{0.400pt}{4.818pt}}
\put(1120,82){\makebox(0,0){1.30}}
\put(1120.0,840.0){\rule[-0.200pt]{0.400pt}{4.818pt}}
\put(1279.0,123.0){\rule[-0.200pt]{0.400pt}{4.818pt}}
\put(1279,82){\makebox(0,0){1.35}}
\put(1279.0,840.0){\rule[-0.200pt]{0.400pt}{4.818pt}}
\put(1439.0,123.0){\rule[-0.200pt]{0.400pt}{4.818pt}}
\put(1439,82){\makebox(0,0){1.40}}
\put(1439.0,840.0){\rule[-0.200pt]{0.400pt}{4.818pt}}
\put(161.0,123.0){\rule[-0.200pt]{307.870pt}{0.400pt}}
\put(1439.0,123.0){\rule[-0.200pt]{0.400pt}{177.543pt}}
\put(161.0,860.0){\rule[-0.200pt]{307.870pt}{0.400pt}}
\put(40,491){\makebox(0,0){$R_{e,N}^2/N\rule{7mm}{0pt}$}}
\put(800,21){\makebox(0,0){$\omega$}}
\put(912,624){\makebox(0,0)[l]{$\omega_\theta$}}
\put(161.0,123.0){\rule[-0.200pt]{0.400pt}{177.543pt}}
\multiput(891.25,622.92)(-1.309,-0.499){115}{\rule{1.144pt}{0.120pt}}
\multiput(893.63,623.17)(-151.625,-59.000){2}{\rule{0.572pt}{0.400pt}}
\put(742,565){\vector(-3,-1){0}}
\put(1279,820){\makebox(0,0)[r]{1024}}
\put(161,695){\circle{12}}
\put(321,663){\circle{12}}
\put(481,627){\circle{12}}
\put(640,584){\circle{12}}
\put(672,574){\circle{12}}
\put(704,564){\circle{12}}
\put(736,554){\circle{12}}
\put(768,543){\circle{12}}
\put(800,531){\circle{12}}
\put(832,518){\circle{12}}
\put(864,504){\circle{12}}
\put(896,489){\circle{12}}
\put(928,472){\circle{12}}
\put(960,453){\circle{12}}
\put(992,429){\circle{12}}
\put(1024,399){\circle{12}}
\put(1056,354){\circle{12}}
\put(1088,287){\circle{12}}
\put(1104,247){\circle{12}}
\put(1120,213){\circle{12}}
\put(1151,174){\circle{12}}
\put(1183,161){\circle{12}}
\put(1215,154){\circle{12}}
\put(1247,150){\circle{12}}
\put(1247,150){\circle{12}}
\put(1279,147){\circle{12}}
\put(1311,145){\circle{12}}
\put(1311,145){\circle{12}}
\put(1343,143){\circle{12}}
\put(1343,143){\circle{12}}
\put(1375,141){\circle{12}}
\put(1407,140){\circle{12}}
\put(1407,140){\circle{12}}
\put(1112,226){\circle{12}}
\put(1349,820){\circle{12}}
\put(1279,779){\makebox(0,0)[r]{2048}}
\put(161,712){\makebox(0,0){$+$}}
\put(321,677){\makebox(0,0){$+$}}
\put(481,637){\makebox(0,0){$+$}}
\put(640,590){\makebox(0,0){$+$}}
\put(672,579){\makebox(0,0){$+$}}
\put(704,567){\makebox(0,0){$+$}}
\put(736,555){\makebox(0,0){$+$}}
\put(768,542){\makebox(0,0){$+$}}
\put(800,527){\makebox(0,0){$+$}}
\put(832,512){\makebox(0,0){$+$}}
\put(864,494){\makebox(0,0){$+$}}
\put(896,474){\makebox(0,0){$+$}}
\put(928,449){\makebox(0,0){$+$}}
\put(960,415){\makebox(0,0){$+$}}
\put(992,344){\makebox(0,0){$+$}}
\put(1008,269){\makebox(0,0){$+$}}
\put(1024,197){\makebox(0,0){$+$}}
\put(1056,156){\makebox(0,0){$+$}}
\put(1088,148){\makebox(0,0){$+$}}
\put(1120,143){\makebox(0,0){$+$}}
\put(1120,143){\makebox(0,0){$+$}}
\put(1151,141){\makebox(0,0){$+$}}
\put(1151,141){\makebox(0,0){$+$}}
\put(1183,138){\makebox(0,0){$+$}}
\put(1015,230){\makebox(0,0){$+$}}
\put(1349,779){\makebox(0,0){$+$}}
\sbox{\plotpoint}{\rule[-0.400pt]{0.800pt}{0.800pt}}
\put(1279,738){\makebox(0,0)[r]{4096}}
\put(161,727){\makebox(0,0){$\star$}}
\put(161,727){\makebox(0,0){$\star$}}
\put(321,691){\makebox(0,0){$\star$}}
\put(481,648){\makebox(0,0){$\star$}}
\put(640,595){\makebox(0,0){$\star$}}
\put(704,569){\makebox(0,0){$\star$}}
\put(736,555){\makebox(0,0){$\star$}}
\put(768,540){\makebox(0,0){$\star$}}
\put(800,523){\makebox(0,0){$\star$}}
\put(832,504){\makebox(0,0){$\star$}}
\put(864,482){\makebox(0,0){$\star$}}
\put(896,454){\makebox(0,0){$\star$}}
\put(928,401){\makebox(0,0){$\star$}}
\put(928,401){\makebox(0,0){$\star$}}
\put(944,287){\makebox(0,0){$\star$}}
\put(944,288){\makebox(0,0){$\star$}}
\put(947,250){\makebox(0,0){$\star$}}
\put(947,248){\makebox(0,0){$\star$}}
\put(950,216){\makebox(0,0){$\star$}}
\put(950,213){\makebox(0,0){$\star$}}
\put(953,189){\makebox(0,0){$\star$}}
\put(953,193){\makebox(0,0){$\star$}}
\put(960,165){\makebox(0,0){$\star$}}
\put(960,163){\makebox(0,0){$\star$}}
\put(960,165){\makebox(0,0){$\star$}}
\put(992,145){\makebox(0,0){$\star$}}
\put(992,144){\makebox(0,0){$\star$}}
\put(1024,139){\makebox(0,0){$\star$}}
\put(1056,136){\makebox(0,0){$\star$}}
\put(1056,136){\makebox(0,0){$\star$}}
\put(948,242){\makebox(0,0){$\star$}}
\put(1349,738){\makebox(0,0){$\star$}}
\sbox{\plotpoint}{\rule[-0.500pt]{1.000pt}{1.000pt}}
\put(1279,697){\makebox(0,0)[r]{8192}}
\put(161,742){\makebox(0,0){$\times$}}
\put(481,657){\makebox(0,0){$\times$}}
\put(736,556){\makebox(0,0){$\times$}}
\put(864,468){\makebox(0,0){$\times$}}
\put(870,460){\makebox(0,0){$\times$}}
\put(883,442){\makebox(0,0){$\times$}}
\put(889,418){\makebox(0,0){$\times$}}
\put(896,361){\makebox(0,0){$\times$}}
\put(897,303){\makebox(0,0){$\times$}}
\put(899,274){\makebox(0,0){$\times$}}
\put(901,233){\makebox(0,0){$\times$}}
\put(902,201){\makebox(0,0){$\times$}}
\put(909,156){\makebox(0,0){$\times$}}
\put(915,143){\makebox(0,0){$\times$}}
\put(921,145){\makebox(0,0){$\times$}}
\put(928,141){\makebox(0,0){$\times$}}
\put(900,265){\makebox(0,0){$\times$}}
\put(1349,697){\makebox(0,0){$\times$}}
\sbox{\plotpoint}{\rule[-0.600pt]{1.200pt}{1.200pt}}
\put(1279,656){\makebox(0,0)[r]{16384}}
\put(856,457){\circle*{12}}
\put(861,405){\circle*{12}}
\put(862,359){\circle*{12}}
\put(864,218){\circle*{12}}
\put(865,173){\circle*{12}}
\put(867,155){\circle*{12}}
\put(869,150){\circle*{12}}
\put(872,145){\circle*{12}}
\put(1349,656){\circle*{12}}
\end{picture}
\caption{\it $R_{e,N}^2/N$ versus $\omega$ for lengths $N=1024$,
$2048$, $4096$, $8192$, and $16384$.} 
\label{figure1} 
\end{center}
\end{figure}
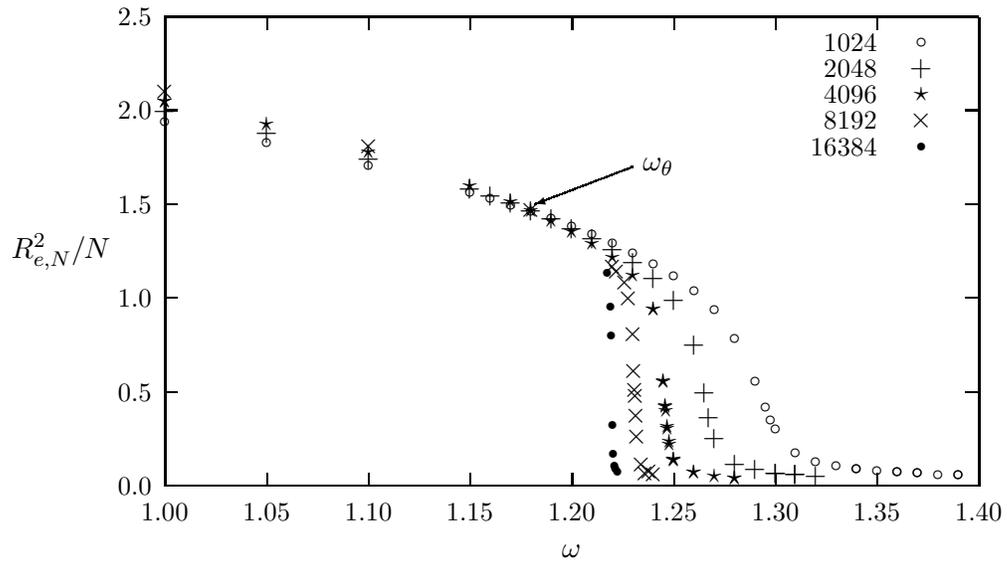

In discussing our findings on the nature of this polymer collapse
transition, it is natural to first present the change of size of the
polymer as the interaction strength changes. In Figure \ref{figure1},
we display the mean-squared end-to-end distance normalised by
walk length, $R_{e,N}^2/N$ for lengths $1024$ up to $16384$ (we have
analogous data for the quantity $R_{m,N}^2/N$). As discussed, one
expects this quantity to increase logarithmically in $N$ in the
swollen regime, approach a constant at a random-walk like
$\theta$-point and decrease in $N$ as a power law in the collapsed
regime. As can be seen clearly from Figure \ref{figure1}, there is
indeed a transition from a swollen region, where this parameter
increases with $N$ to a collapsed region, where the value has dropped
sharply. In the transition region, however, two phenomena can be
noticed. On the one hand, around $\omega=1.18$ the quantity
$R_{e,N}^2/N$ approaches a constant, which is indicative of
$\theta$-point behaviour.  On the other hand, the collapse occurs in a
region which is well separated from this $\theta$-point. With
increasing polymer length, the region where the collapse occurs
approaches the $\theta$-point, but simultaneously sharpens so strongly
that it remains well separated from it.

\begin{figure}[p]
\begin{center}
\setlength{\unitlength}{0.240900pt}
\ifx\plotpoint\undefined\newsavebox{\plotpoint}\fi
\sbox{\plotpoint}{\rule[-0.200pt]{0.400pt}{0.400pt}}
\begin{picture}(1500,900)(0,0)
\font\gnuplot=cmr10 at 10pt
\gnuplot
\sbox{\plotpoint}{\rule[-0.200pt]{0.400pt}{0.400pt}}
\put(161.0,123.0){\rule[-0.200pt]{4.818pt}{0.400pt}}
\put(141,123){\makebox(0,0)[r]{1.5}}
\put(1419.0,123.0){\rule[-0.200pt]{4.818pt}{0.400pt}}
\put(161.0,228.0){\rule[-0.200pt]{4.818pt}{0.400pt}}
\put(141,228){\makebox(0,0)[r]{1.6}}
\put(1419.0,228.0){\rule[-0.200pt]{4.818pt}{0.400pt}}
\put(161.0,334.0){\rule[-0.200pt]{4.818pt}{0.400pt}}
\put(141,334){\makebox(0,0)[r]{1.7}}
\put(1419.0,334.0){\rule[-0.200pt]{4.818pt}{0.400pt}}
\put(161.0,439.0){\rule[-0.200pt]{4.818pt}{0.400pt}}
\put(141,439){\makebox(0,0)[r]{1.8}}
\put(1419.0,439.0){\rule[-0.200pt]{4.818pt}{0.400pt}}
\put(161.0,544.0){\rule[-0.200pt]{4.818pt}{0.400pt}}
\put(141,544){\makebox(0,0)[r]{1.9}}
\put(1419.0,544.0){\rule[-0.200pt]{4.818pt}{0.400pt}}
\put(161.0,649.0){\rule[-0.200pt]{4.818pt}{0.400pt}}
\put(141,649){\makebox(0,0)[r]{2.0}}
\put(1419.0,649.0){\rule[-0.200pt]{4.818pt}{0.400pt}}
\put(161.0,755.0){\rule[-0.200pt]{4.818pt}{0.400pt}}
\put(141,755){\makebox(0,0)[r]{2.1}}
\put(1419.0,755.0){\rule[-0.200pt]{4.818pt}{0.400pt}}
\put(161.0,860.0){\rule[-0.200pt]{4.818pt}{0.400pt}}
\put(141,860){\makebox(0,0)[r]{2.2}}
\put(1419.0,860.0){\rule[-0.200pt]{4.818pt}{0.400pt}}
\put(161.0,123.0){\rule[-0.200pt]{0.400pt}{4.818pt}}
\put(161,82){\makebox(0,0){100}}
\put(161.0,840.0){\rule[-0.200pt]{0.400pt}{4.818pt}}
\put(353.0,123.0){\rule[-0.200pt]{0.400pt}{2.409pt}}
\put(353.0,850.0){\rule[-0.200pt]{0.400pt}{2.409pt}}
\put(466.0,123.0){\rule[-0.200pt]{0.400pt}{2.409pt}}
\put(466.0,850.0){\rule[-0.200pt]{0.400pt}{2.409pt}}
\put(546.0,123.0){\rule[-0.200pt]{0.400pt}{2.409pt}}
\put(546.0,850.0){\rule[-0.200pt]{0.400pt}{2.409pt}}
\put(608.0,123.0){\rule[-0.200pt]{0.400pt}{2.409pt}}
\put(608.0,850.0){\rule[-0.200pt]{0.400pt}{2.409pt}}
\put(658.0,123.0){\rule[-0.200pt]{0.400pt}{2.409pt}}
\put(658.0,850.0){\rule[-0.200pt]{0.400pt}{2.409pt}}
\put(701.0,123.0){\rule[-0.200pt]{0.400pt}{2.409pt}}
\put(701.0,850.0){\rule[-0.200pt]{0.400pt}{2.409pt}}
\put(738.0,123.0){\rule[-0.200pt]{0.400pt}{2.409pt}}
\put(738.0,850.0){\rule[-0.200pt]{0.400pt}{2.409pt}}
\put(771.0,123.0){\rule[-0.200pt]{0.400pt}{2.409pt}}
\put(771.0,850.0){\rule[-0.200pt]{0.400pt}{2.409pt}}
\put(800.0,123.0){\rule[-0.200pt]{0.400pt}{4.818pt}}
\put(800,82){\makebox(0,0){1000}}
\put(800.0,840.0){\rule[-0.200pt]{0.400pt}{4.818pt}}
\put(992.0,123.0){\rule[-0.200pt]{0.400pt}{2.409pt}}
\put(992.0,850.0){\rule[-0.200pt]{0.400pt}{2.409pt}}
\put(1105.0,123.0){\rule[-0.200pt]{0.400pt}{2.409pt}}
\put(1105.0,850.0){\rule[-0.200pt]{0.400pt}{2.409pt}}
\put(1185.0,123.0){\rule[-0.200pt]{0.400pt}{2.409pt}}
\put(1185.0,850.0){\rule[-0.200pt]{0.400pt}{2.409pt}}
\put(1247.0,123.0){\rule[-0.200pt]{0.400pt}{2.409pt}}
\put(1247.0,850.0){\rule[-0.200pt]{0.400pt}{2.409pt}}
\put(1297.0,123.0){\rule[-0.200pt]{0.400pt}{2.409pt}}
\put(1297.0,850.0){\rule[-0.200pt]{0.400pt}{2.409pt}}
\put(1340.0,123.0){\rule[-0.200pt]{0.400pt}{2.409pt}}
\put(1340.0,850.0){\rule[-0.200pt]{0.400pt}{2.409pt}}
\put(1377.0,123.0){\rule[-0.200pt]{0.400pt}{2.409pt}}
\put(1377.0,850.0){\rule[-0.200pt]{0.400pt}{2.409pt}}
\put(1410.0,123.0){\rule[-0.200pt]{0.400pt}{2.409pt}}
\put(1410.0,850.0){\rule[-0.200pt]{0.400pt}{2.409pt}}
\put(1439.0,123.0){\rule[-0.200pt]{0.400pt}{4.818pt}}
\put(1439,82){\makebox(0,0){10000}}
\put(1439.0,840.0){\rule[-0.200pt]{0.400pt}{4.818pt}}
\put(161.0,123.0){\rule[-0.200pt]{307.870pt}{0.400pt}}
\put(1439.0,123.0){\rule[-0.200pt]{0.400pt}{177.543pt}}
\put(161.0,860.0){\rule[-0.200pt]{307.870pt}{0.400pt}}
\put(40,491){\makebox(0,0){$R_{e,N}^2/N\rule{7mm}{0pt}$}}
\put(800,21){\makebox(0,0){$N$}}
\put(800,702){\makebox(0,0)[l]{$\omega=1.0$}}
\put(1105,334){\makebox(0,0)[l]{$\omega=1.1$}}
\put(161.0,123.0){\rule[-0.200pt]{0.400pt}{177.543pt}}
\put(230.0,368.0){\rule[-0.200pt]{0.400pt}{1.686pt}}
\put(220.0,368.0){\rule[-0.200pt]{4.818pt}{0.400pt}}
\put(220.0,375.0){\rule[-0.200pt]{4.818pt}{0.400pt}}
\put(326.0,408.0){\rule[-0.200pt]{0.400pt}{1.686pt}}
\put(316.0,408.0){\rule[-0.200pt]{4.818pt}{0.400pt}}
\put(316.0,415.0){\rule[-0.200pt]{4.818pt}{0.400pt}}
\put(422.0,445.0){\rule[-0.200pt]{0.400pt}{1.927pt}}
\put(412.0,445.0){\rule[-0.200pt]{4.818pt}{0.400pt}}
\put(412.0,453.0){\rule[-0.200pt]{4.818pt}{0.400pt}}
\put(518.0,481.0){\rule[-0.200pt]{0.400pt}{1.927pt}}
\put(508.0,481.0){\rule[-0.200pt]{4.818pt}{0.400pt}}
\put(508.0,489.0){\rule[-0.200pt]{4.818pt}{0.400pt}}
\put(614.0,516.0){\rule[-0.200pt]{0.400pt}{1.927pt}}
\put(604.0,516.0){\rule[-0.200pt]{4.818pt}{0.400pt}}
\put(604.0,524.0){\rule[-0.200pt]{4.818pt}{0.400pt}}
\put(710.0,551.0){\rule[-0.200pt]{0.400pt}{1.686pt}}
\put(700.0,551.0){\rule[-0.200pt]{4.818pt}{0.400pt}}
\put(700.0,558.0){\rule[-0.200pt]{4.818pt}{0.400pt}}
\put(807.0,583.0){\rule[-0.200pt]{0.400pt}{1.927pt}}
\put(797.0,583.0){\rule[-0.200pt]{4.818pt}{0.400pt}}
\put(797.0,591.0){\rule[-0.200pt]{4.818pt}{0.400pt}}
\put(903.0,612.0){\rule[-0.200pt]{0.400pt}{1.927pt}}
\put(893.0,612.0){\rule[-0.200pt]{4.818pt}{0.400pt}}
\put(893.0,620.0){\rule[-0.200pt]{4.818pt}{0.400pt}}
\put(999.0,642.0){\rule[-0.200pt]{0.400pt}{1.927pt}}
\put(989.0,642.0){\rule[-0.200pt]{4.818pt}{0.400pt}}
\put(989.0,650.0){\rule[-0.200pt]{4.818pt}{0.400pt}}
\put(1095.0,672.0){\rule[-0.200pt]{0.400pt}{1.927pt}}
\put(1085.0,672.0){\rule[-0.200pt]{4.818pt}{0.400pt}}
\put(1085.0,680.0){\rule[-0.200pt]{4.818pt}{0.400pt}}
\put(1191.0,701.0){\rule[-0.200pt]{0.400pt}{2.168pt}}
\put(1181.0,701.0){\rule[-0.200pt]{4.818pt}{0.400pt}}
\put(1181.0,710.0){\rule[-0.200pt]{4.818pt}{0.400pt}}
\put(1287.0,727.0){\rule[-0.200pt]{0.400pt}{1.927pt}}
\put(1277.0,727.0){\rule[-0.200pt]{4.818pt}{0.400pt}}
\put(1277.0,735.0){\rule[-0.200pt]{4.818pt}{0.400pt}}
\put(1384.0,751.0){\rule[-0.200pt]{0.400pt}{2.168pt}}
\put(1374.0,751.0){\rule[-0.200pt]{4.818pt}{0.400pt}}
\put(230,371){\rule{1pt}{1pt}}
\put(326,411){\rule{1pt}{1pt}}
\put(422,449){\rule{1pt}{1pt}}
\put(518,485){\rule{1pt}{1pt}}
\put(614,520){\rule{1pt}{1pt}}
\put(710,554){\rule{1pt}{1pt}}
\put(807,587){\rule{1pt}{1pt}}
\put(903,616){\rule{1pt}{1pt}}
\put(999,646){\rule{1pt}{1pt}}
\put(1095,676){\rule{1pt}{1pt}}
\put(1191,705){\rule{1pt}{1pt}}
\put(1287,731){\rule{1pt}{1pt}}
\put(1384,755){\rule{1pt}{1pt}}
\put(1374.0,760.0){\rule[-0.200pt]{4.818pt}{0.400pt}}
\put(161,341){\usebox{\plotpoint}}
\put(161.00,341.00){\usebox{\plotpoint}}
\put(179.66,350.08){\usebox{\plotpoint}}
\put(198.72,358.30){\usebox{\plotpoint}}
\put(217.75,366.58){\usebox{\plotpoint}}
\put(236.76,374.88){\usebox{\plotpoint}}
\put(255.69,383.34){\usebox{\plotpoint}}
\put(274.93,391.13){\usebox{\plotpoint}}
\put(294.20,398.83){\usebox{\plotpoint}}
\put(313.02,407.54){\usebox{\plotpoint}}
\put(332.27,415.28){\usebox{\plotpoint}}
\put(351.49,423.11){\usebox{\plotpoint}}
\put(370.73,430.89){\usebox{\plotpoint}}
\put(390.30,437.81){\usebox{\plotpoint}}
\put(409.52,445.63){\usebox{\plotpoint}}
\put(428.77,453.37){\usebox{\plotpoint}}
\put(448.30,460.38){\usebox{\plotpoint}}
\put(467.58,468.07){\usebox{\plotpoint}}
\put(486.90,475.63){\usebox{\plotpoint}}
\put(506.38,482.76){\usebox{\plotpoint}}
\put(525.95,489.67){\usebox{\plotpoint}}
\put(545.50,496.62){\usebox{\plotpoint}}
\put(564.96,503.76){\usebox{\plotpoint}}
\put(584.35,511.12){\usebox{\plotpoint}}
\put(603.86,518.18){\usebox{\plotpoint}}
\put(623.52,524.84){\usebox{\plotpoint}}
\put(643.00,532.00){\usebox{\plotpoint}}
\put(662.69,538.56){\usebox{\plotpoint}}
\put(682.16,545.72){\usebox{\plotpoint}}
\put(701.85,552.28){\usebox{\plotpoint}}
\put(721.64,558.55){\usebox{\plotpoint}}
\put(741.00,566.00){\usebox{\plotpoint}}
\put(760.78,572.26){\usebox{\plotpoint}}
\put(780.57,578.52){\usebox{\plotpoint}}
\put(800.26,585.09){\usebox{\plotpoint}}
\put(820.05,591.35){\usebox{\plotpoint}}
\put(839.74,597.91){\usebox{\plotpoint}}
\put(859.53,604.18){\usebox{\plotpoint}}
\put(879.23,610.69){\usebox{\plotpoint}}
\put(899.00,617.00){\usebox{\plotpoint}}
\put(918.73,623.46){\usebox{\plotpoint}}
\put(938.74,628.94){\usebox{\plotpoint}}
\put(958.49,635.31){\usebox{\plotpoint}}
\put(978.22,641.74){\usebox{\plotpoint}}
\put(997.99,648.07){\usebox{\plotpoint}}
\put(1018.00,653.54){\usebox{\plotpoint}}
\put(1037.75,659.92){\usebox{\plotpoint}}
\put(1057.77,665.39){\usebox{\plotpoint}}
\put(1077.50,671.83){\usebox{\plotpoint}}
\put(1097.46,677.41){\usebox{\plotpoint}}
\put(1117.34,683.33){\usebox{\plotpoint}}
\put(1137.27,689.09){\usebox{\plotpoint}}
\put(1157.14,695.04){\usebox{\plotpoint}}
\put(1177.08,700.77){\usebox{\plotpoint}}
\put(1196.97,706.66){\usebox{\plotpoint}}
\put(1216.89,712.47){\usebox{\plotpoint}}
\put(1236.76,718.41){\usebox{\plotpoint}}
\put(1256.75,723.92){\usebox{\plotpoint}}
\put(1276.87,728.97){\usebox{\plotpoint}}
\put(1296.74,734.91){\usebox{\plotpoint}}
\put(1316.77,740.26){\usebox{\plotpoint}}
\put(1336.75,745.84){\usebox{\plotpoint}}
\put(1356.73,751.43){\usebox{\plotpoint}}
\put(1376.78,756.78){\usebox{\plotpoint}}
\put(1384,759){\usebox{\plotpoint}}
\sbox{\plotpoint}{\rule[-0.400pt]{0.800pt}{0.800pt}}
\put(230.0,208.0){\usebox{\plotpoint}}
\put(220.0,208.0){\rule[-0.400pt]{4.818pt}{0.800pt}}
\put(220.0,211.0){\rule[-0.400pt]{4.818pt}{0.800pt}}
\put(326.0,233.0){\usebox{\plotpoint}}
\put(316.0,233.0){\rule[-0.400pt]{4.818pt}{0.800pt}}
\put(316.0,236.0){\rule[-0.400pt]{4.818pt}{0.800pt}}
\put(422.0,256.0){\rule[-0.400pt]{0.800pt}{0.964pt}}
\put(412.0,256.0){\rule[-0.400pt]{4.818pt}{0.800pt}}
\put(412.0,260.0){\rule[-0.400pt]{4.818pt}{0.800pt}}
\put(518.0,279.0){\usebox{\plotpoint}}
\put(508.0,279.0){\rule[-0.400pt]{4.818pt}{0.800pt}}
\put(508.0,282.0){\rule[-0.400pt]{4.818pt}{0.800pt}}
\put(614.0,300.0){\usebox{\plotpoint}}
\put(604.0,300.0){\rule[-0.400pt]{4.818pt}{0.800pt}}
\put(604.0,303.0){\rule[-0.400pt]{4.818pt}{0.800pt}}
\put(710.0,321.0){\rule[-0.400pt]{0.800pt}{0.964pt}}
\put(700.0,321.0){\rule[-0.400pt]{4.818pt}{0.800pt}}
\put(700.0,325.0){\rule[-0.400pt]{4.818pt}{0.800pt}}
\put(807.0,342.0){\usebox{\plotpoint}}
\put(797.0,342.0){\rule[-0.400pt]{4.818pt}{0.800pt}}
\put(797.0,345.0){\rule[-0.400pt]{4.818pt}{0.800pt}}
\put(903.0,362.0){\usebox{\plotpoint}}
\put(893.0,362.0){\rule[-0.400pt]{4.818pt}{0.800pt}}
\put(893.0,365.0){\rule[-0.400pt]{4.818pt}{0.800pt}}
\put(999.0,380.0){\usebox{\plotpoint}}
\put(989.0,380.0){\rule[-0.400pt]{4.818pt}{0.800pt}}
\put(989.0,383.0){\rule[-0.400pt]{4.818pt}{0.800pt}}
\put(1095.0,400.0){\usebox{\plotpoint}}
\put(1085.0,400.0){\rule[-0.400pt]{4.818pt}{0.800pt}}
\put(1085.0,403.0){\rule[-0.400pt]{4.818pt}{0.800pt}}
\put(1191.0,416.0){\rule[-0.400pt]{0.800pt}{0.964pt}}
\put(1181.0,416.0){\rule[-0.400pt]{4.818pt}{0.800pt}}
\put(1181.0,420.0){\rule[-0.400pt]{4.818pt}{0.800pt}}
\put(1287.0,434.0){\rule[-0.400pt]{0.800pt}{0.964pt}}
\put(1277.0,434.0){\rule[-0.400pt]{4.818pt}{0.800pt}}
\put(1277.0,438.0){\rule[-0.400pt]{4.818pt}{0.800pt}}
\put(1384.0,451.0){\usebox{\plotpoint}}
\put(1374.0,451.0){\rule[-0.400pt]{4.818pt}{0.800pt}}
\put(230,210){\rule{1pt}{1pt}}
\put(326,234){\rule{1pt}{1pt}}
\put(422,258){\rule{1pt}{1pt}}
\put(518,280){\rule{1pt}{1pt}}
\put(614,301){\rule{1pt}{1pt}}
\put(710,323){\rule{1pt}{1pt}}
\put(807,343){\rule{1pt}{1pt}}
\put(903,363){\rule{1pt}{1pt}}
\put(999,382){\rule{1pt}{1pt}}
\put(1095,401){\rule{1pt}{1pt}}
\put(1191,418){\rule{1pt}{1pt}}
\put(1287,436){\rule{1pt}{1pt}}
\put(1384,453){\rule{1pt}{1pt}}
\put(1374.0,454.0){\rule[-0.400pt]{4.818pt}{0.800pt}}
\sbox{\plotpoint}{\rule[-0.200pt]{0.400pt}{0.400pt}}
\put(161,193){\usebox{\plotpoint}}
\put(161.00,193.00){\usebox{\plotpoint}}
\put(181.17,197.89){\usebox{\plotpoint}}
\put(201.06,203.76){\usebox{\plotpoint}}
\put(221.24,208.59){\usebox{\plotpoint}}
\put(241.39,213.60){\usebox{\plotpoint}}
\put(261.58,218.39){\usebox{\plotpoint}}
\put(281.76,223.25){\usebox{\plotpoint}}
\put(301.63,229.16){\usebox{\plotpoint}}
\put(321.83,233.96){\usebox{\plotpoint}}
\put(341.96,238.99){\usebox{\plotpoint}}
\put(362.16,243.79){\usebox{\plotpoint}}
\put(382.29,248.82){\usebox{\plotpoint}}
\put(402.48,253.62){\usebox{\plotpoint}}
\put(422.63,258.61){\usebox{\plotpoint}}
\put(442.81,263.45){\usebox{\plotpoint}}
\put(463.06,267.93){\usebox{\plotpoint}}
\put(483.32,272.33){\usebox{\plotpoint}}
\put(503.50,277.19){\usebox{\plotpoint}}
\put(523.67,282.08){\usebox{\plotpoint}}
\put(543.84,286.96){\usebox{\plotpoint}}
\put(564.13,291.25){\usebox{\plotpoint}}
\put(584.36,295.84){\usebox{\plotpoint}}
\put(604.54,300.66){\usebox{\plotpoint}}
\put(624.80,305.13){\usebox{\plotpoint}}
\put(645.07,309.52){\usebox{\plotpoint}}
\put(665.38,313.73){\usebox{\plotpoint}}
\put(685.60,318.40){\usebox{\plotpoint}}
\put(705.77,323.27){\usebox{\plotpoint}}
\put(726.11,327.28){\usebox{\plotpoint}}
\put(746.35,331.82){\usebox{\plotpoint}}
\put(766.63,336.10){\usebox{\plotpoint}}
\put(787.00,340.00){\usebox{\plotpoint}}
\put(807.21,344.65){\usebox{\plotpoint}}
\put(827.51,348.88){\usebox{\plotpoint}}
\put(847.88,352.82){\usebox{\plotpoint}}
\put(868.23,356.81){\usebox{\plotpoint}}
\put(888.58,360.78){\usebox{\plotpoint}}
\put(908.86,365.14){\usebox{\plotpoint}}
\put(929.17,369.36){\usebox{\plotpoint}}
\put(949.47,373.62){\usebox{\plotpoint}}
\put(969.84,377.46){\usebox{\plotpoint}}
\put(990.19,381.50){\usebox{\plotpoint}}
\put(1010.53,385.59){\usebox{\plotpoint}}
\put(1030.94,389.24){\usebox{\plotpoint}}
\put(1051.30,393.20){\usebox{\plotpoint}}
\put(1071.65,397.16){\usebox{\plotpoint}}
\put(1092.08,400.78){\usebox{\plotpoint}}
\put(1112.37,405.06){\usebox{\plotpoint}}
\put(1132.86,408.36){\usebox{\plotpoint}}
\put(1153.14,412.69){\usebox{\plotpoint}}
\put(1173.64,415.94){\usebox{\plotpoint}}
\put(1193.91,420.32){\usebox{\plotpoint}}
\put(1214.41,423.57){\usebox{\plotpoint}}
\put(1234.69,427.92){\usebox{\plotpoint}}
\put(1255.18,431.20){\usebox{\plotpoint}}
\put(1275.68,434.45){\usebox{\plotpoint}}
\put(1295.97,438.74){\usebox{\plotpoint}}
\put(1316.45,442.07){\usebox{\plotpoint}}
\put(1336.93,445.45){\usebox{\plotpoint}}
\put(1357.42,448.74){\usebox{\plotpoint}}
\put(1377.71,453.03){\usebox{\plotpoint}}
\put(1384,454){\usebox{\plotpoint}}
\end{picture}
\caption{\it $R_{e,N}^2/N$ versus $N$ at $\omega=1.0$ and
$\omega=1.1$. The curves are fits to $R_{e,N}^2/N=a\log(N+b)^c$ 
over the shown range.}
\label{figure2} 
\end{center}
\end{figure}
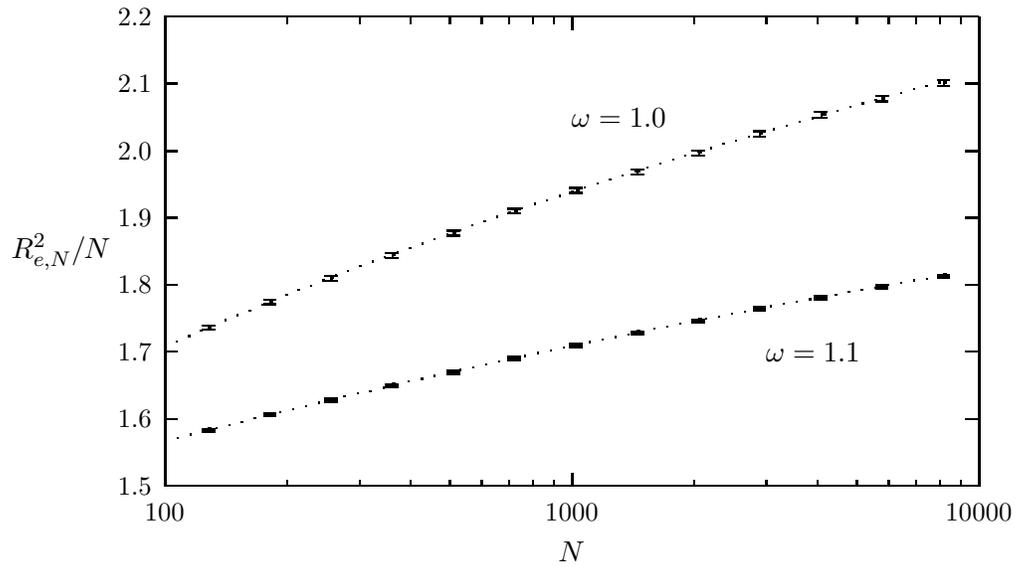

In the swollen phase, our results are in correspondence with the
logarithmic corrections seen by Grassberger
\cite{grassberger1994a-a}, see Figure \ref{figure2}. As in that paper,
we observe that $R_{e,N}^2$ grows as $N(\log N)^c$, albeit not with the
exponent predicted by field theory.  At $\omega=1.0$ we find
$c=0.30$. This value shifts to $c=0.22$ at $\omega=1.1$, indicating
the presence of strong temperature-dependent correction terms.

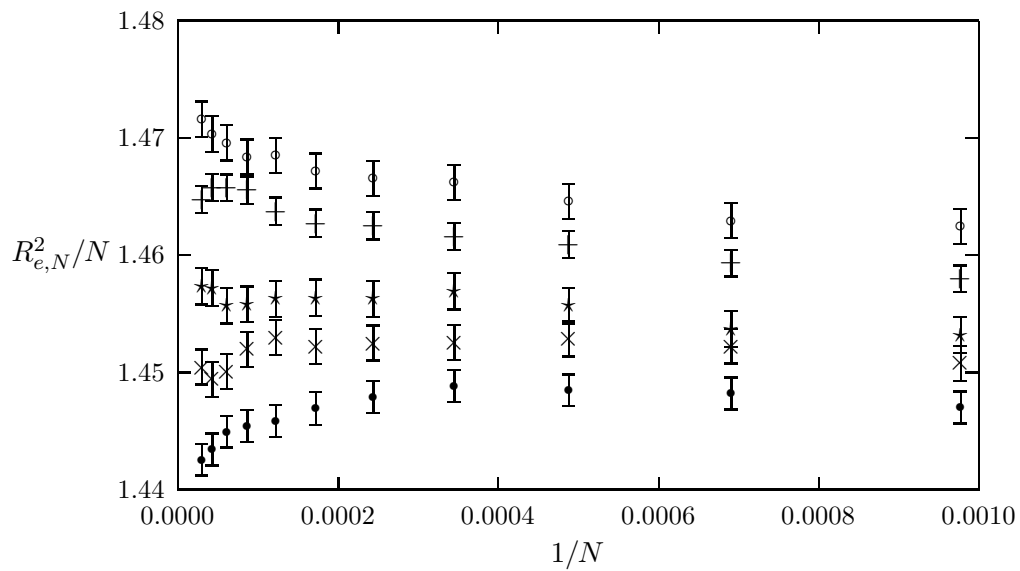
\begin{figure}[p]
\begin{center}
\setlength{\unitlength}{0.240900pt}
\ifx\plotpoint\undefined\newsavebox{\plotpoint}\fi
\sbox{\plotpoint}{\rule[-0.200pt]{0.400pt}{0.400pt}}
\begin{picture}(1500,900)(0,0)
\font\gnuplot=cmr10 at 10pt
\gnuplot
\sbox{\plotpoint}{\rule[-0.200pt]{0.400pt}{0.400pt}}
\put(181.0,123.0){\rule[-0.200pt]{4.818pt}{0.400pt}}
\put(161,123){\makebox(0,0)[r]{1.44}}
\put(1419.0,123.0){\rule[-0.200pt]{4.818pt}{0.400pt}}
\put(181.0,307.0){\rule[-0.200pt]{4.818pt}{0.400pt}}
\put(161,307){\makebox(0,0)[r]{1.45}}
\put(1419.0,307.0){\rule[-0.200pt]{4.818pt}{0.400pt}}
\put(181.0,492.0){\rule[-0.200pt]{4.818pt}{0.400pt}}
\put(161,492){\makebox(0,0)[r]{1.46}}
\put(1419.0,492.0){\rule[-0.200pt]{4.818pt}{0.400pt}}
\put(181.0,676.0){\rule[-0.200pt]{4.818pt}{0.400pt}}
\put(161,676){\makebox(0,0)[r]{1.47}}
\put(1419.0,676.0){\rule[-0.200pt]{4.818pt}{0.400pt}}
\put(181.0,860.0){\rule[-0.200pt]{4.818pt}{0.400pt}}
\put(161,860){\makebox(0,0)[r]{1.48}}
\put(1419.0,860.0){\rule[-0.200pt]{4.818pt}{0.400pt}}
\put(181.0,123.0){\rule[-0.200pt]{0.400pt}{4.818pt}}
\put(181,82){\makebox(0,0){0.0000}}
\put(181.0,840.0){\rule[-0.200pt]{0.400pt}{4.818pt}}
\put(433.0,123.0){\rule[-0.200pt]{0.400pt}{4.818pt}}
\put(433,82){\makebox(0,0){0.0002}}
\put(433.0,840.0){\rule[-0.200pt]{0.400pt}{4.818pt}}
\put(684.0,123.0){\rule[-0.200pt]{0.400pt}{4.818pt}}
\put(684,82){\makebox(0,0){0.0004}}
\put(684.0,840.0){\rule[-0.200pt]{0.400pt}{4.818pt}}
\put(936.0,123.0){\rule[-0.200pt]{0.400pt}{4.818pt}}
\put(936,82){\makebox(0,0){0.0006}}
\put(936.0,840.0){\rule[-0.200pt]{0.400pt}{4.818pt}}
\put(1187.0,123.0){\rule[-0.200pt]{0.400pt}{4.818pt}}
\put(1187,82){\makebox(0,0){0.0008}}
\put(1187.0,840.0){\rule[-0.200pt]{0.400pt}{4.818pt}}
\put(1439.0,123.0){\rule[-0.200pt]{0.400pt}{4.818pt}}
\put(1439,82){\makebox(0,0){0.0010}}
\put(1439.0,840.0){\rule[-0.200pt]{0.400pt}{4.818pt}}
\put(181.0,123.0){\rule[-0.200pt]{303.052pt}{0.400pt}}
\put(1439.0,123.0){\rule[-0.200pt]{0.400pt}{177.543pt}}
\put(181.0,860.0){\rule[-0.200pt]{303.052pt}{0.400pt}}
\put(40,491){\makebox(0,0){$R_{e,N}^2/N\rule{7mm}{0pt}$}}
\put(810,21){\makebox(0,0){$1/N$}}
\put(181.0,123.0){\rule[-0.200pt]{0.400pt}{177.543pt}}
\put(1410.0,509.0){\rule[-0.200pt]{0.400pt}{13.249pt}}
\put(1400.0,509.0){\rule[-0.200pt]{4.818pt}{0.400pt}}
\put(1400.0,564.0){\rule[-0.200pt]{4.818pt}{0.400pt}}
\put(1050.0,518.0){\rule[-0.200pt]{0.400pt}{13.249pt}}
\put(1040.0,518.0){\rule[-0.200pt]{4.818pt}{0.400pt}}
\put(1040.0,573.0){\rule[-0.200pt]{4.818pt}{0.400pt}}
\put(795.0,548.0){\rule[-0.200pt]{0.400pt}{13.249pt}}
\put(785.0,548.0){\rule[-0.200pt]{4.818pt}{0.400pt}}
\put(785.0,603.0){\rule[-0.200pt]{4.818pt}{0.400pt}}
\put(615.0,578.0){\rule[-0.200pt]{0.400pt}{13.249pt}}
\put(605.0,578.0){\rule[-0.200pt]{4.818pt}{0.400pt}}
\put(605.0,633.0){\rule[-0.200pt]{4.818pt}{0.400pt}}
\put(488.0,584.0){\rule[-0.200pt]{0.400pt}{13.249pt}}
\put(478.0,584.0){\rule[-0.200pt]{4.818pt}{0.400pt}}
\put(478.0,639.0){\rule[-0.200pt]{4.818pt}{0.400pt}}
\put(398.0,596.0){\rule[-0.200pt]{0.400pt}{13.249pt}}
\put(388.0,596.0){\rule[-0.200pt]{4.818pt}{0.400pt}}
\put(388.0,651.0){\rule[-0.200pt]{4.818pt}{0.400pt}}
\put(335.0,620.0){\rule[-0.200pt]{0.400pt}{13.490pt}}
\put(325.0,620.0){\rule[-0.200pt]{4.818pt}{0.400pt}}
\put(325.0,676.0){\rule[-0.200pt]{4.818pt}{0.400pt}}
\put(290.0,618.0){\rule[-0.200pt]{0.400pt}{13.249pt}}
\put(280.0,618.0){\rule[-0.200pt]{4.818pt}{0.400pt}}
\put(280.0,673.0){\rule[-0.200pt]{4.818pt}{0.400pt}}
\put(258.0,640.0){\rule[-0.200pt]{0.400pt}{13.490pt}}
\put(248.0,640.0){\rule[-0.200pt]{4.818pt}{0.400pt}}
\put(248.0,696.0){\rule[-0.200pt]{4.818pt}{0.400pt}}
\put(235.0,654.0){\rule[-0.200pt]{0.400pt}{13.490pt}}
\put(225.0,654.0){\rule[-0.200pt]{4.818pt}{0.400pt}}
\put(225.0,710.0){\rule[-0.200pt]{4.818pt}{0.400pt}}
\put(219.0,677.0){\rule[-0.200pt]{0.400pt}{13.490pt}}
\put(209.0,677.0){\rule[-0.200pt]{4.818pt}{0.400pt}}
\put(1410,537){\circle{12}}
\put(1050,545){\circle{12}}
\put(795,576){\circle{12}}
\put(615,606){\circle{12}}
\put(488,612){\circle{12}}
\put(398,624){\circle{12}}
\put(335,648){\circle{12}}
\put(290,646){\circle{12}}
\put(258,668){\circle{12}}
\put(235,682){\circle{12}}
\put(219,705){\circle{12}}
\put(209.0,733.0){\rule[-0.200pt]{4.818pt}{0.400pt}}
\put(1410.0,433.0){\rule[-0.200pt]{0.400pt}{10.118pt}}
\put(1400.0,433.0){\rule[-0.200pt]{4.818pt}{0.400pt}}
\put(1400.0,475.0){\rule[-0.200pt]{4.818pt}{0.400pt}}
\put(1050.0,458.0){\rule[-0.200pt]{0.400pt}{10.118pt}}
\put(1040.0,458.0){\rule[-0.200pt]{4.818pt}{0.400pt}}
\put(1040.0,500.0){\rule[-0.200pt]{4.818pt}{0.400pt}}
\put(795.0,487.0){\rule[-0.200pt]{0.400pt}{10.118pt}}
\put(785.0,487.0){\rule[-0.200pt]{4.818pt}{0.400pt}}
\put(785.0,529.0){\rule[-0.200pt]{4.818pt}{0.400pt}}
\put(615.0,499.0){\rule[-0.200pt]{0.400pt}{10.359pt}}
\put(605.0,499.0){\rule[-0.200pt]{4.818pt}{0.400pt}}
\put(605.0,542.0){\rule[-0.200pt]{4.818pt}{0.400pt}}
\put(488.0,516.0){\rule[-0.200pt]{0.400pt}{10.359pt}}
\put(478.0,516.0){\rule[-0.200pt]{4.818pt}{0.400pt}}
\put(478.0,559.0){\rule[-0.200pt]{4.818pt}{0.400pt}}
\put(398.0,520.0){\rule[-0.200pt]{0.400pt}{10.359pt}}
\put(388.0,520.0){\rule[-0.200pt]{4.818pt}{0.400pt}}
\put(388.0,563.0){\rule[-0.200pt]{4.818pt}{0.400pt}}
\put(335.0,539.0){\rule[-0.200pt]{0.400pt}{10.359pt}}
\put(325.0,539.0){\rule[-0.200pt]{4.818pt}{0.400pt}}
\put(325.0,582.0){\rule[-0.200pt]{4.818pt}{0.400pt}}
\put(290.0,572.0){\rule[-0.200pt]{0.400pt}{10.359pt}}
\put(280.0,572.0){\rule[-0.200pt]{4.818pt}{0.400pt}}
\put(280.0,615.0){\rule[-0.200pt]{4.818pt}{0.400pt}}
\put(258.0,576.0){\rule[-0.200pt]{0.400pt}{10.118pt}}
\put(248.0,576.0){\rule[-0.200pt]{4.818pt}{0.400pt}}
\put(248.0,618.0){\rule[-0.200pt]{4.818pt}{0.400pt}}
\put(235.0,577.0){\rule[-0.200pt]{0.400pt}{10.118pt}}
\put(225.0,577.0){\rule[-0.200pt]{4.818pt}{0.400pt}}
\put(225.0,619.0){\rule[-0.200pt]{4.818pt}{0.400pt}}
\put(219.0,558.0){\rule[-0.200pt]{0.400pt}{10.118pt}}
\put(209.0,558.0){\rule[-0.200pt]{4.818pt}{0.400pt}}
\put(1410,454){\makebox(0,0){$+$}}
\put(1050,479){\makebox(0,0){$+$}}
\put(795,508){\makebox(0,0){$+$}}
\put(615,520){\makebox(0,0){$+$}}
\put(488,537){\makebox(0,0){$+$}}
\put(398,541){\makebox(0,0){$+$}}
\put(335,560){\makebox(0,0){$+$}}
\put(290,594){\makebox(0,0){$+$}}
\put(258,597){\makebox(0,0){$+$}}
\put(235,598){\makebox(0,0){$+$}}
\put(219,579){\makebox(0,0){$+$}}
\put(209.0,600.0){\rule[-0.200pt]{4.818pt}{0.400pt}}
\put(1410.0,338.0){\rule[-0.200pt]{0.400pt}{13.490pt}}
\put(1400.0,338.0){\rule[-0.200pt]{4.818pt}{0.400pt}}
\put(1400.0,394.0){\rule[-0.200pt]{4.818pt}{0.400pt}}
\put(1050.0,347.0){\rule[-0.200pt]{0.400pt}{13.731pt}}
\put(1040.0,347.0){\rule[-0.200pt]{4.818pt}{0.400pt}}
\put(1040.0,404.0){\rule[-0.200pt]{4.818pt}{0.400pt}}
\put(795.0,383.0){\rule[-0.200pt]{0.400pt}{13.731pt}}
\put(785.0,383.0){\rule[-0.200pt]{4.818pt}{0.400pt}}
\put(785.0,440.0){\rule[-0.200pt]{4.818pt}{0.400pt}}
\put(615.0,406.0){\rule[-0.200pt]{0.400pt}{13.731pt}}
\put(605.0,406.0){\rule[-0.200pt]{4.818pt}{0.400pt}}
\put(605.0,463.0){\rule[-0.200pt]{4.818pt}{0.400pt}}
\put(488.0,394.0){\rule[-0.200pt]{0.400pt}{13.731pt}}
\put(478.0,394.0){\rule[-0.200pt]{4.818pt}{0.400pt}}
\put(478.0,451.0){\rule[-0.200pt]{4.818pt}{0.400pt}}
\put(398.0,396.0){\rule[-0.200pt]{0.400pt}{13.731pt}}
\put(388.0,396.0){\rule[-0.200pt]{4.818pt}{0.400pt}}
\put(388.0,453.0){\rule[-0.200pt]{4.818pt}{0.400pt}}
\put(335.0,394.0){\rule[-0.200pt]{0.400pt}{13.731pt}}
\put(325.0,394.0){\rule[-0.200pt]{4.818pt}{0.400pt}}
\put(325.0,451.0){\rule[-0.200pt]{4.818pt}{0.400pt}}
\put(290.0,386.0){\rule[-0.200pt]{0.400pt}{13.490pt}}
\put(280.0,386.0){\rule[-0.200pt]{4.818pt}{0.400pt}}
\put(280.0,442.0){\rule[-0.200pt]{4.818pt}{0.400pt}}
\put(258.0,384.0){\rule[-0.200pt]{0.400pt}{13.490pt}}
\put(248.0,384.0){\rule[-0.200pt]{4.818pt}{0.400pt}}
\put(248.0,440.0){\rule[-0.200pt]{4.818pt}{0.400pt}}
\put(235.0,412.0){\rule[-0.200pt]{0.400pt}{13.490pt}}
\put(225.0,412.0){\rule[-0.200pt]{4.818pt}{0.400pt}}
\put(225.0,468.0){\rule[-0.200pt]{4.818pt}{0.400pt}}
\put(219.0,414.0){\rule[-0.200pt]{0.400pt}{13.731pt}}
\put(209.0,414.0){\rule[-0.200pt]{4.818pt}{0.400pt}}
\put(1410,366){\makebox(0,0){$\star$}}
\put(1050,375){\makebox(0,0){$\star$}}
\put(795,412){\makebox(0,0){$\star$}}
\put(615,434){\makebox(0,0){$\star$}}
\put(488,423){\makebox(0,0){$\star$}}
\put(398,424){\makebox(0,0){$\star$}}
\put(335,423){\makebox(0,0){$\star$}}
\put(290,414){\makebox(0,0){$\star$}}
\put(258,412){\makebox(0,0){$\star$}}
\put(235,440){\makebox(0,0){$\star$}}
\put(219,443){\makebox(0,0){$\star$}}
\put(209.0,471.0){\rule[-0.200pt]{4.818pt}{0.400pt}}
\put(1410.0,294.0){\rule[-0.200pt]{0.400pt}{13.249pt}}
\put(1400.0,294.0){\rule[-0.200pt]{4.818pt}{0.400pt}}
\put(1400.0,349.0){\rule[-0.200pt]{4.818pt}{0.400pt}}
\put(1050.0,321.0){\rule[-0.200pt]{0.400pt}{13.009pt}}
\put(1040.0,321.0){\rule[-0.200pt]{4.818pt}{0.400pt}}
\put(1040.0,375.0){\rule[-0.200pt]{4.818pt}{0.400pt}}
\put(795.0,332.0){\rule[-0.200pt]{0.400pt}{13.249pt}}
\put(785.0,332.0){\rule[-0.200pt]{4.818pt}{0.400pt}}
\put(785.0,387.0){\rule[-0.200pt]{4.818pt}{0.400pt}}
\put(615.0,327.0){\rule[-0.200pt]{0.400pt}{13.249pt}}
\put(605.0,327.0){\rule[-0.200pt]{4.818pt}{0.400pt}}
\put(605.0,382.0){\rule[-0.200pt]{4.818pt}{0.400pt}}
\put(488.0,326.0){\rule[-0.200pt]{0.400pt}{13.249pt}}
\put(478.0,326.0){\rule[-0.200pt]{4.818pt}{0.400pt}}
\put(478.0,381.0){\rule[-0.200pt]{4.818pt}{0.400pt}}
\put(398.0,320.0){\rule[-0.200pt]{0.400pt}{13.249pt}}
\put(388.0,320.0){\rule[-0.200pt]{4.818pt}{0.400pt}}
\put(388.0,375.0){\rule[-0.200pt]{4.818pt}{0.400pt}}
\put(335.0,334.0){\rule[-0.200pt]{0.400pt}{13.249pt}}
\put(325.0,334.0){\rule[-0.200pt]{4.818pt}{0.400pt}}
\put(325.0,389.0){\rule[-0.200pt]{4.818pt}{0.400pt}}
\put(290.0,316.0){\rule[-0.200pt]{0.400pt}{13.249pt}}
\put(280.0,316.0){\rule[-0.200pt]{4.818pt}{0.400pt}}
\put(280.0,371.0){\rule[-0.200pt]{4.818pt}{0.400pt}}
\put(258.0,281.0){\rule[-0.200pt]{0.400pt}{13.249pt}}
\put(248.0,281.0){\rule[-0.200pt]{4.818pt}{0.400pt}}
\put(248.0,336.0){\rule[-0.200pt]{4.818pt}{0.400pt}}
\put(235.0,269.0){\rule[-0.200pt]{0.400pt}{13.249pt}}
\put(225.0,269.0){\rule[-0.200pt]{4.818pt}{0.400pt}}
\put(225.0,324.0){\rule[-0.200pt]{4.818pt}{0.400pt}}
\put(219.0,288.0){\rule[-0.200pt]{0.400pt}{13.249pt}}
\put(209.0,288.0){\rule[-0.200pt]{4.818pt}{0.400pt}}
\put(1410,322){\makebox(0,0){$\times$}}
\put(1050,348){\makebox(0,0){$\times$}}
\put(795,360){\makebox(0,0){$\times$}}
\put(615,354){\makebox(0,0){$\times$}}
\put(488,353){\makebox(0,0){$\times$}}
\put(398,348){\makebox(0,0){$\times$}}
\put(335,361){\makebox(0,0){$\times$}}
\put(290,344){\makebox(0,0){$\times$}}
\put(258,309){\makebox(0,0){$\times$}}
\put(235,297){\makebox(0,0){$\times$}}
\put(219,315){\makebox(0,0){$\times$}}
\put(209.0,343.0){\rule[-0.200pt]{4.818pt}{0.400pt}}
\put(1410.0,227.0){\rule[-0.200pt]{0.400pt}{12.045pt}}
\put(1400.0,227.0){\rule[-0.200pt]{4.818pt}{0.400pt}}
\put(1400.0,277.0){\rule[-0.200pt]{4.818pt}{0.400pt}}
\put(1050.0,249.0){\rule[-0.200pt]{0.400pt}{12.045pt}}
\put(1040.0,249.0){\rule[-0.200pt]{4.818pt}{0.400pt}}
\put(1040.0,299.0){\rule[-0.200pt]{4.818pt}{0.400pt}}
\put(795.0,254.0){\rule[-0.200pt]{0.400pt}{12.045pt}}
\put(785.0,254.0){\rule[-0.200pt]{4.818pt}{0.400pt}}
\put(785.0,304.0){\rule[-0.200pt]{4.818pt}{0.400pt}}
\put(615.0,261.0){\rule[-0.200pt]{0.400pt}{12.045pt}}
\put(605.0,261.0){\rule[-0.200pt]{4.818pt}{0.400pt}}
\put(605.0,311.0){\rule[-0.200pt]{4.818pt}{0.400pt}}
\put(488.0,243.0){\rule[-0.200pt]{0.400pt}{12.286pt}}
\put(478.0,243.0){\rule[-0.200pt]{4.818pt}{0.400pt}}
\put(478.0,294.0){\rule[-0.200pt]{4.818pt}{0.400pt}}
\put(398.0,225.0){\rule[-0.200pt]{0.400pt}{12.286pt}}
\put(388.0,225.0){\rule[-0.200pt]{4.818pt}{0.400pt}}
\put(388.0,276.0){\rule[-0.200pt]{4.818pt}{0.400pt}}
\put(335.0,206.0){\rule[-0.200pt]{0.400pt}{12.045pt}}
\put(325.0,206.0){\rule[-0.200pt]{4.818pt}{0.400pt}}
\put(325.0,256.0){\rule[-0.200pt]{4.818pt}{0.400pt}}
\put(290.0,198.0){\rule[-0.200pt]{0.400pt}{12.045pt}}
\put(280.0,198.0){\rule[-0.200pt]{4.818pt}{0.400pt}}
\put(280.0,248.0){\rule[-0.200pt]{4.818pt}{0.400pt}}
\put(258.0,189.0){\rule[-0.200pt]{0.400pt}{12.045pt}}
\put(248.0,189.0){\rule[-0.200pt]{4.818pt}{0.400pt}}
\put(248.0,239.0){\rule[-0.200pt]{4.818pt}{0.400pt}}
\put(235.0,161.0){\rule[-0.200pt]{0.400pt}{12.045pt}}
\put(225.0,161.0){\rule[-0.200pt]{4.818pt}{0.400pt}}
\put(225.0,211.0){\rule[-0.200pt]{4.818pt}{0.400pt}}
\put(219.0,145.0){\rule[-0.200pt]{0.400pt}{12.045pt}}
\put(209.0,145.0){\rule[-0.200pt]{4.818pt}{0.400pt}}
\put(1410,252){\circle*{12}}
\put(1050,274){\circle*{12}}
\put(795,279){\circle*{12}}
\put(615,286){\circle*{12}}
\put(488,269){\circle*{12}}
\put(398,251){\circle*{12}}
\put(335,231){\circle*{12}}
\put(290,223){\circle*{12}}
\put(258,214){\circle*{12}}
\put(235,186){\circle*{12}}
\put(219,170){\circle*{12}}
\put(209.0,195.0){\rule[-0.200pt]{4.818pt}{0.400pt}}
\end{picture}
\caption{\it $R_{e,N}^2/N$ versus $1/N$ in the $\theta$-region:
$\omega=1.180$, $1.181$, $1.182$, $1.183$, $1.184$ from top to bottom.} 
\label{figure3} 
\end{center}
\end{figure}

Near the suspected $\theta$-point, we extended our simulations to
walks of length $32768$. Figure \ref{figure3} shows a plot of
$R_{e,N}^2/N$ versus $1/N$ for values of $\omega$ between $1.180$ and
$1.184$. At $\omega=1.182$ we have indeed an asymptotically linear
dependence of $R_{e,N}^2$ on $N$. Moreover, at $\omega=1.182$ we
estimate from our data $B_N=R_{m,N}^2/R_{e,N}^2=0.500(1)$, which is
also indicative of Gaussian behaviour.

\begin{figure}[p]
\begin{center}
\setlength{\unitlength}{0.240900pt}
\ifx\plotpoint\undefined\newsavebox{\plotpoint}\fi
\sbox{\plotpoint}{\rule[-0.200pt]{0.400pt}{0.400pt}}
\begin{picture}(1500,900)(0,0)
\font\gnuplot=cmr10 at 10pt
\gnuplot
\sbox{\plotpoint}{\rule[-0.200pt]{0.400pt}{0.400pt}}
\put(161.0,123.0){\rule[-0.200pt]{4.818pt}{0.400pt}}
\put(141,123){\makebox(0,0)[r]{0}}
\put(1419.0,123.0){\rule[-0.200pt]{4.818pt}{0.400pt}}
\put(161.0,228.0){\rule[-0.200pt]{4.818pt}{0.400pt}}
\put(141,228){\makebox(0,0)[r]{20}}
\put(1419.0,228.0){\rule[-0.200pt]{4.818pt}{0.400pt}}
\put(161.0,334.0){\rule[-0.200pt]{4.818pt}{0.400pt}}
\put(141,334){\makebox(0,0)[r]{40}}
\put(1419.0,334.0){\rule[-0.200pt]{4.818pt}{0.400pt}}
\put(161.0,439.0){\rule[-0.200pt]{4.818pt}{0.400pt}}
\put(141,439){\makebox(0,0)[r]{60}}
\put(1419.0,439.0){\rule[-0.200pt]{4.818pt}{0.400pt}}
\put(161.0,544.0){\rule[-0.200pt]{4.818pt}{0.400pt}}
\put(141,544){\makebox(0,0)[r]{80}}
\put(1419.0,544.0){\rule[-0.200pt]{4.818pt}{0.400pt}}
\put(161.0,649.0){\rule[-0.200pt]{4.818pt}{0.400pt}}
\put(141,649){\makebox(0,0)[r]{100}}
\put(1419.0,649.0){\rule[-0.200pt]{4.818pt}{0.400pt}}
\put(161.0,755.0){\rule[-0.200pt]{4.818pt}{0.400pt}}
\put(141,755){\makebox(0,0)[r]{120}}
\put(1419.0,755.0){\rule[-0.200pt]{4.818pt}{0.400pt}}
\put(161.0,860.0){\rule[-0.200pt]{4.818pt}{0.400pt}}
\put(141,860){\makebox(0,0)[r]{140}}
\put(1419.0,860.0){\rule[-0.200pt]{4.818pt}{0.400pt}}
\put(161.0,123.0){\rule[-0.200pt]{0.400pt}{4.818pt}}
\put(161,82){\makebox(0,0){1}}
\put(161.0,840.0){\rule[-0.200pt]{0.400pt}{4.818pt}}
\put(257.0,123.0){\rule[-0.200pt]{0.400pt}{2.409pt}}
\put(257.0,850.0){\rule[-0.200pt]{0.400pt}{2.409pt}}
\put(313.0,123.0){\rule[-0.200pt]{0.400pt}{2.409pt}}
\put(313.0,850.0){\rule[-0.200pt]{0.400pt}{2.409pt}}
\put(353.0,123.0){\rule[-0.200pt]{0.400pt}{2.409pt}}
\put(353.0,850.0){\rule[-0.200pt]{0.400pt}{2.409pt}}
\put(384.0,123.0){\rule[-0.200pt]{0.400pt}{2.409pt}}
\put(384.0,850.0){\rule[-0.200pt]{0.400pt}{2.409pt}}
\put(410.0,123.0){\rule[-0.200pt]{0.400pt}{2.409pt}}
\put(410.0,850.0){\rule[-0.200pt]{0.400pt}{2.409pt}}
\put(431.0,123.0){\rule[-0.200pt]{0.400pt}{2.409pt}}
\put(431.0,850.0){\rule[-0.200pt]{0.400pt}{2.409pt}}
\put(450.0,123.0){\rule[-0.200pt]{0.400pt}{2.409pt}}
\put(450.0,850.0){\rule[-0.200pt]{0.400pt}{2.409pt}}
\put(466.0,123.0){\rule[-0.200pt]{0.400pt}{2.409pt}}
\put(466.0,850.0){\rule[-0.200pt]{0.400pt}{2.409pt}}
\put(481.0,123.0){\rule[-0.200pt]{0.400pt}{4.818pt}}
\put(481,82){\makebox(0,0){10}}
\put(481.0,840.0){\rule[-0.200pt]{0.400pt}{4.818pt}}
\put(577.0,123.0){\rule[-0.200pt]{0.400pt}{2.409pt}}
\put(577.0,850.0){\rule[-0.200pt]{0.400pt}{2.409pt}}
\put(633.0,123.0){\rule[-0.200pt]{0.400pt}{2.409pt}}
\put(633.0,850.0){\rule[-0.200pt]{0.400pt}{2.409pt}}
\put(673.0,123.0){\rule[-0.200pt]{0.400pt}{2.409pt}}
\put(673.0,850.0){\rule[-0.200pt]{0.400pt}{2.409pt}}
\put(704.0,123.0){\rule[-0.200pt]{0.400pt}{2.409pt}}
\put(704.0,850.0){\rule[-0.200pt]{0.400pt}{2.409pt}}
\put(729.0,123.0){\rule[-0.200pt]{0.400pt}{2.409pt}}
\put(729.0,850.0){\rule[-0.200pt]{0.400pt}{2.409pt}}
\put(751.0,123.0){\rule[-0.200pt]{0.400pt}{2.409pt}}
\put(751.0,850.0){\rule[-0.200pt]{0.400pt}{2.409pt}}
\put(769.0,123.0){\rule[-0.200pt]{0.400pt}{2.409pt}}
\put(769.0,850.0){\rule[-0.200pt]{0.400pt}{2.409pt}}
\put(785.0,123.0){\rule[-0.200pt]{0.400pt}{2.409pt}}
\put(785.0,850.0){\rule[-0.200pt]{0.400pt}{2.409pt}}
\put(800.0,123.0){\rule[-0.200pt]{0.400pt}{4.818pt}}
\put(800,82){\makebox(0,0){100}}
\put(800.0,840.0){\rule[-0.200pt]{0.400pt}{4.818pt}}
\put(896.0,123.0){\rule[-0.200pt]{0.400pt}{2.409pt}}
\put(896.0,850.0){\rule[-0.200pt]{0.400pt}{2.409pt}}
\put(952.0,123.0){\rule[-0.200pt]{0.400pt}{2.409pt}}
\put(952.0,850.0){\rule[-0.200pt]{0.400pt}{2.409pt}}
\put(992.0,123.0){\rule[-0.200pt]{0.400pt}{2.409pt}}
\put(992.0,850.0){\rule[-0.200pt]{0.400pt}{2.409pt}}
\put(1023.0,123.0){\rule[-0.200pt]{0.400pt}{2.409pt}}
\put(1023.0,850.0){\rule[-0.200pt]{0.400pt}{2.409pt}}
\put(1049.0,123.0){\rule[-0.200pt]{0.400pt}{2.409pt}}
\put(1049.0,850.0){\rule[-0.200pt]{0.400pt}{2.409pt}}
\put(1070.0,123.0){\rule[-0.200pt]{0.400pt}{2.409pt}}
\put(1070.0,850.0){\rule[-0.200pt]{0.400pt}{2.409pt}}
\put(1089.0,123.0){\rule[-0.200pt]{0.400pt}{2.409pt}}
\put(1089.0,850.0){\rule[-0.200pt]{0.400pt}{2.409pt}}
\put(1105.0,123.0){\rule[-0.200pt]{0.400pt}{2.409pt}}
\put(1105.0,850.0){\rule[-0.200pt]{0.400pt}{2.409pt}}
\put(1120.0,123.0){\rule[-0.200pt]{0.400pt}{4.818pt}}
\put(1120,82){\makebox(0,0){1000}}
\put(1120.0,840.0){\rule[-0.200pt]{0.400pt}{4.818pt}}
\put(1216.0,123.0){\rule[-0.200pt]{0.400pt}{2.409pt}}
\put(1216.0,850.0){\rule[-0.200pt]{0.400pt}{2.409pt}}
\put(1272.0,123.0){\rule[-0.200pt]{0.400pt}{2.409pt}}
\put(1272.0,850.0){\rule[-0.200pt]{0.400pt}{2.409pt}}
\put(1312.0,123.0){\rule[-0.200pt]{0.400pt}{2.409pt}}
\put(1312.0,850.0){\rule[-0.200pt]{0.400pt}{2.409pt}}
\put(1343.0,123.0){\rule[-0.200pt]{0.400pt}{2.409pt}}
\put(1343.0,850.0){\rule[-0.200pt]{0.400pt}{2.409pt}}
\put(1368.0,123.0){\rule[-0.200pt]{0.400pt}{2.409pt}}
\put(1368.0,850.0){\rule[-0.200pt]{0.400pt}{2.409pt}}
\put(1390.0,123.0){\rule[-0.200pt]{0.400pt}{2.409pt}}
\put(1390.0,850.0){\rule[-0.200pt]{0.400pt}{2.409pt}}
\put(1408.0,123.0){\rule[-0.200pt]{0.400pt}{2.409pt}}
\put(1408.0,850.0){\rule[-0.200pt]{0.400pt}{2.409pt}}
\put(1424.0,123.0){\rule[-0.200pt]{0.400pt}{2.409pt}}
\put(1424.0,850.0){\rule[-0.200pt]{0.400pt}{2.409pt}}
\put(1439.0,123.0){\rule[-0.200pt]{0.400pt}{4.818pt}}
\put(1439,82){\makebox(0,0){10000}}
\put(1439.0,840.0){\rule[-0.200pt]{0.400pt}{4.818pt}}
\put(161.0,123.0){\rule[-0.200pt]{307.870pt}{0.400pt}}
\put(1439.0,123.0){\rule[-0.200pt]{0.400pt}{177.543pt}}
\put(161.0,860.0){\rule[-0.200pt]{307.870pt}{0.400pt}}
\put(40,491){\makebox(0,0){$R_{e,N}^2\rule{7mm}{0pt}$}}
\put(800,21){\makebox(0,0){$N$}}
\put(161.0,123.0){\rule[-0.200pt]{0.400pt}{177.543pt}}
\put(257,135){\usebox{\plotpoint}}
\put(247.0,135.0){\rule[-0.200pt]{4.818pt}{0.400pt}}
\put(247.0,135.0){\rule[-0.200pt]{4.818pt}{0.400pt}}
\put(353.0,148.0){\usebox{\plotpoint}}
\put(343.0,148.0){\rule[-0.200pt]{4.818pt}{0.400pt}}
\put(343.0,149.0){\rule[-0.200pt]{4.818pt}{0.400pt}}
\put(384.0,154.0){\usebox{\plotpoint}}
\put(374.0,154.0){\rule[-0.200pt]{4.818pt}{0.400pt}}
\put(374.0,155.0){\rule[-0.200pt]{4.818pt}{0.400pt}}
\put(450.0,174.0){\usebox{\plotpoint}}
\put(440.0,174.0){\rule[-0.200pt]{4.818pt}{0.400pt}}
\put(440.0,175.0){\rule[-0.200pt]{4.818pt}{0.400pt}}
\put(494.0,193.0){\rule[-0.200pt]{0.400pt}{0.482pt}}
\put(484.0,193.0){\rule[-0.200pt]{4.818pt}{0.400pt}}
\put(484.0,195.0){\rule[-0.200pt]{4.818pt}{0.400pt}}
\put(546.0,224.0){\rule[-0.200pt]{0.400pt}{0.723pt}}
\put(536.0,224.0){\rule[-0.200pt]{4.818pt}{0.400pt}}
\put(536.0,227.0){\rule[-0.200pt]{4.818pt}{0.400pt}}
\put(590.0,259.0){\rule[-0.200pt]{0.400pt}{0.964pt}}
\put(580.0,259.0){\rule[-0.200pt]{4.818pt}{0.400pt}}
\put(580.0,263.0){\rule[-0.200pt]{4.818pt}{0.400pt}}
\put(642.0,315.0){\rule[-0.200pt]{0.400pt}{1.445pt}}
\put(632.0,315.0){\rule[-0.200pt]{4.818pt}{0.400pt}}
\put(632.0,321.0){\rule[-0.200pt]{4.818pt}{0.400pt}}
\put(689.0,382.0){\rule[-0.200pt]{0.400pt}{1.927pt}}
\put(679.0,382.0){\rule[-0.200pt]{4.818pt}{0.400pt}}
\put(679.0,390.0){\rule[-0.200pt]{4.818pt}{0.400pt}}
\put(738.0,470.0){\rule[-0.200pt]{0.400pt}{2.650pt}}
\put(728.0,470.0){\rule[-0.200pt]{4.818pt}{0.400pt}}
\put(728.0,481.0){\rule[-0.200pt]{4.818pt}{0.400pt}}
\put(785.0,572.0){\rule[-0.200pt]{0.400pt}{3.613pt}}
\put(775.0,572.0){\rule[-0.200pt]{4.818pt}{0.400pt}}
\put(775.0,587.0){\rule[-0.200pt]{4.818pt}{0.400pt}}
\put(834.0,685.0){\rule[-0.200pt]{0.400pt}{4.577pt}}
\put(824.0,685.0){\rule[-0.200pt]{4.818pt}{0.400pt}}
\put(824.0,704.0){\rule[-0.200pt]{4.818pt}{0.400pt}}
\put(882.0,762.0){\rule[-0.200pt]{0.400pt}{5.782pt}}
\put(872.0,762.0){\rule[-0.200pt]{4.818pt}{0.400pt}}
\put(872.0,786.0){\rule[-0.200pt]{4.818pt}{0.400pt}}
\put(930.0,710.0){\rule[-0.200pt]{0.400pt}{6.263pt}}
\put(920.0,710.0){\rule[-0.200pt]{4.818pt}{0.400pt}}
\put(920.0,736.0){\rule[-0.200pt]{4.818pt}{0.400pt}}
\put(979.0,528.0){\rule[-0.200pt]{0.400pt}{4.336pt}}
\put(969.0,528.0){\rule[-0.200pt]{4.818pt}{0.400pt}}
\put(969.0,546.0){\rule[-0.200pt]{4.818pt}{0.400pt}}
\put(1027.0,443.0){\rule[-0.200pt]{0.400pt}{2.891pt}}
\put(1017.0,443.0){\rule[-0.200pt]{4.818pt}{0.400pt}}
\put(1017.0,455.0){\rule[-0.200pt]{4.818pt}{0.400pt}}
\put(1075.0,423.0){\rule[-0.200pt]{0.400pt}{2.650pt}}
\put(1065.0,423.0){\rule[-0.200pt]{4.818pt}{0.400pt}}
\put(1065.0,434.0){\rule[-0.200pt]{4.818pt}{0.400pt}}
\put(1123.0,426.0){\rule[-0.200pt]{0.400pt}{2.409pt}}
\put(1113.0,426.0){\rule[-0.200pt]{4.818pt}{0.400pt}}
\put(257,135){\circle*{12}}
\put(353,148){\circle*{12}}
\put(384,155){\circle*{12}}
\put(450,175){\circle*{12}}
\put(494,194){\circle*{12}}
\put(546,225){\circle*{12}}
\put(590,261){\circle*{12}}
\put(642,318){\circle*{12}}
\put(689,386){\circle*{12}}
\put(738,476){\circle*{12}}
\put(785,580){\circle*{12}}
\put(834,694){\circle*{12}}
\put(882,774){\circle*{12}}
\put(930,723){\circle*{12}}
\put(979,537){\circle*{12}}
\put(1027,449){\circle*{12}}
\put(1075,429){\circle*{12}}
\put(1123,431){\circle*{12}}
\put(1113.0,436.0){\rule[-0.200pt]{4.818pt}{0.400pt}}
\end{picture}
\caption{\it $R_{e,N}^2$ versus $N$ for $\omega=1.4$.}
\label{figure4} 
\end{center}
\end{figure}
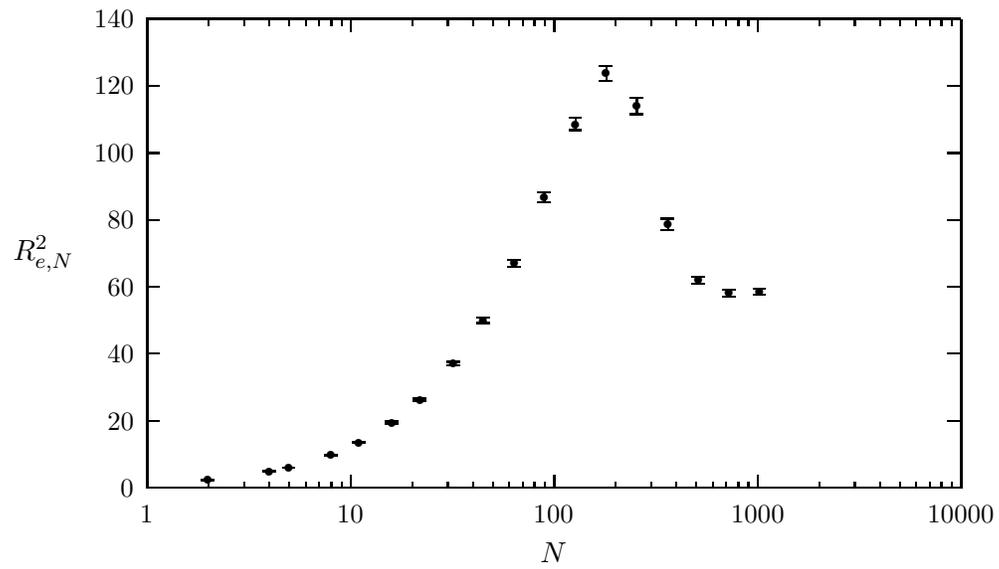

As was already seen from Figure \ref{figure1}, the collapse happens
very rapidly as $\omega$ increases.  An alternative way of visualising
this is to consider how the size of the polymer changes at fixed
$\omega$ in the collapsed phase as the length $N$ increases. As shown
in Figure \ref{figure4} for $\omega=1.4$, $R_{e,N}^2$ changes
non-monotonically in $N$! After an initial increase, the size of the
polymer actually shrinks around $N=250$ as it undergoes
collapse\footnote{Note that this lack of monotonicity is {\em not} an
indication of a first-order transition. A similar non-monotonous
behaviour can be observed for ISAW in three dimensions, where the
collapse transition is second-order \cite{grassberger1995a-a}.}
corresponding to a rapid increase of the density. For large enough
$N$, we expect to see the true collapsed behaviour, i.e.\ $R_{e,N}^2$
growing again as $N^{1/2}$, but this regime is beyond the reach of our
PERM simulations on current computer hardware.

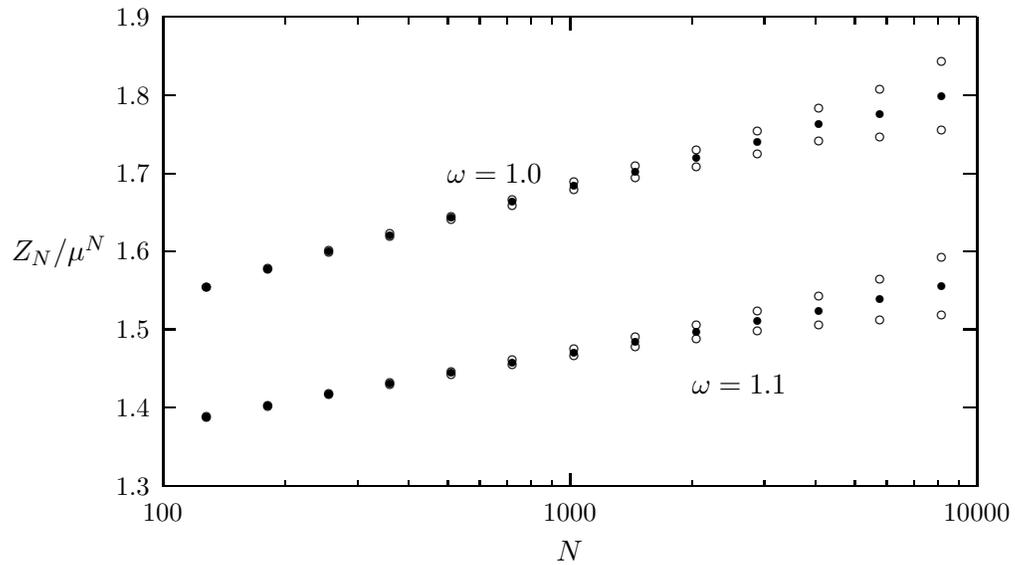
\begin{figure}[p]
\begin{center}
\setlength{\unitlength}{0.240900pt}
\ifx\plotpoint\undefined\newsavebox{\plotpoint}\fi
\sbox{\plotpoint}{\rule[-0.200pt]{0.400pt}{0.400pt}}
\begin{picture}(1500,900)(0,0)
\font\gnuplot=cmr10 at 10pt
\gnuplot
\sbox{\plotpoint}{\rule[-0.200pt]{0.400pt}{0.400pt}}
\put(161.0,123.0){\rule[-0.200pt]{4.818pt}{0.400pt}}
\put(141,123){\makebox(0,0)[r]{1.3}}
\put(1419.0,123.0){\rule[-0.200pt]{4.818pt}{0.400pt}}
\put(161.0,246.0){\rule[-0.200pt]{4.818pt}{0.400pt}}
\put(141,246){\makebox(0,0)[r]{1.4}}
\put(1419.0,246.0){\rule[-0.200pt]{4.818pt}{0.400pt}}
\put(161.0,369.0){\rule[-0.200pt]{4.818pt}{0.400pt}}
\put(141,369){\makebox(0,0)[r]{1.5}}
\put(1419.0,369.0){\rule[-0.200pt]{4.818pt}{0.400pt}}
\put(161.0,492.0){\rule[-0.200pt]{4.818pt}{0.400pt}}
\put(141,492){\makebox(0,0)[r]{1.6}}
\put(1419.0,492.0){\rule[-0.200pt]{4.818pt}{0.400pt}}
\put(161.0,614.0){\rule[-0.200pt]{4.818pt}{0.400pt}}
\put(141,614){\makebox(0,0)[r]{1.7}}
\put(1419.0,614.0){\rule[-0.200pt]{4.818pt}{0.400pt}}
\put(161.0,737.0){\rule[-0.200pt]{4.818pt}{0.400pt}}
\put(141,737){\makebox(0,0)[r]{1.8}}
\put(1419.0,737.0){\rule[-0.200pt]{4.818pt}{0.400pt}}
\put(161.0,860.0){\rule[-0.200pt]{4.818pt}{0.400pt}}
\put(141,860){\makebox(0,0)[r]{1.9}}
\put(1419.0,860.0){\rule[-0.200pt]{4.818pt}{0.400pt}}
\put(161.0,123.0){\rule[-0.200pt]{0.400pt}{4.818pt}}
\put(161,82){\makebox(0,0){100}}
\put(161.0,840.0){\rule[-0.200pt]{0.400pt}{4.818pt}}
\put(353.0,123.0){\rule[-0.200pt]{0.400pt}{2.409pt}}
\put(353.0,850.0){\rule[-0.200pt]{0.400pt}{2.409pt}}
\put(466.0,123.0){\rule[-0.200pt]{0.400pt}{2.409pt}}
\put(466.0,850.0){\rule[-0.200pt]{0.400pt}{2.409pt}}
\put(546.0,123.0){\rule[-0.200pt]{0.400pt}{2.409pt}}
\put(546.0,850.0){\rule[-0.200pt]{0.400pt}{2.409pt}}
\put(608.0,123.0){\rule[-0.200pt]{0.400pt}{2.409pt}}
\put(608.0,850.0){\rule[-0.200pt]{0.400pt}{2.409pt}}
\put(658.0,123.0){\rule[-0.200pt]{0.400pt}{2.409pt}}
\put(658.0,850.0){\rule[-0.200pt]{0.400pt}{2.409pt}}
\put(701.0,123.0){\rule[-0.200pt]{0.400pt}{2.409pt}}
\put(701.0,850.0){\rule[-0.200pt]{0.400pt}{2.409pt}}
\put(738.0,123.0){\rule[-0.200pt]{0.400pt}{2.409pt}}
\put(738.0,850.0){\rule[-0.200pt]{0.400pt}{2.409pt}}
\put(771.0,123.0){\rule[-0.200pt]{0.400pt}{2.409pt}}
\put(771.0,850.0){\rule[-0.200pt]{0.400pt}{2.409pt}}
\put(800.0,123.0){\rule[-0.200pt]{0.400pt}{4.818pt}}
\put(800,82){\makebox(0,0){1000}}
\put(800.0,840.0){\rule[-0.200pt]{0.400pt}{4.818pt}}
\put(992.0,123.0){\rule[-0.200pt]{0.400pt}{2.409pt}}
\put(992.0,850.0){\rule[-0.200pt]{0.400pt}{2.409pt}}
\put(1105.0,123.0){\rule[-0.200pt]{0.400pt}{2.409pt}}
\put(1105.0,850.0){\rule[-0.200pt]{0.400pt}{2.409pt}}
\put(1185.0,123.0){\rule[-0.200pt]{0.400pt}{2.409pt}}
\put(1185.0,850.0){\rule[-0.200pt]{0.400pt}{2.409pt}}
\put(1247.0,123.0){\rule[-0.200pt]{0.400pt}{2.409pt}}
\put(1247.0,850.0){\rule[-0.200pt]{0.400pt}{2.409pt}}
\put(1297.0,123.0){\rule[-0.200pt]{0.400pt}{2.409pt}}
\put(1297.0,850.0){\rule[-0.200pt]{0.400pt}{2.409pt}}
\put(1340.0,123.0){\rule[-0.200pt]{0.400pt}{2.409pt}}
\put(1340.0,850.0){\rule[-0.200pt]{0.400pt}{2.409pt}}
\put(1377.0,123.0){\rule[-0.200pt]{0.400pt}{2.409pt}}
\put(1377.0,850.0){\rule[-0.200pt]{0.400pt}{2.409pt}}
\put(1410.0,123.0){\rule[-0.200pt]{0.400pt}{2.409pt}}
\put(1410.0,850.0){\rule[-0.200pt]{0.400pt}{2.409pt}}
\put(1439.0,123.0){\rule[-0.200pt]{0.400pt}{4.818pt}}
\put(1439,82){\makebox(0,0){10000}}
\put(1439.0,840.0){\rule[-0.200pt]{0.400pt}{4.818pt}}
\put(161.0,123.0){\rule[-0.200pt]{307.870pt}{0.400pt}}
\put(1439.0,123.0){\rule[-0.200pt]{0.400pt}{177.543pt}}
\put(161.0,860.0){\rule[-0.200pt]{307.870pt}{0.400pt}}
\put(40,491){\makebox(0,0){$Z_N/\mu^N\rule{7mm}{0pt}$}}
\put(800,21){\makebox(0,0){$N$}}
\put(608,614){\makebox(0,0)[l]{$\omega=1.0$}}
\put(992,283){\makebox(0,0)[l]{$\omega=1.1$}}
\put(161.0,123.0){\rule[-0.200pt]{0.400pt}{177.543pt}}
\put(230,436){\circle*{12}}
\put(326,464){\circle*{12}}
\put(422,492){\circle*{12}}
\put(518,517){\circle*{12}}
\put(614,544){\circle*{12}}
\put(710,569){\circle*{12}}
\put(807,595){\circle*{12}}
\put(903,617){\circle*{12}}
\put(999,638){\circle*{12}}
\put(1095,663){\circle*{12}}
\put(1191,691){\circle*{12}}
\put(1287,708){\circle*{12}}
\put(1384,736){\circle*{12}}
\put(230,436){\circle{12}}
\put(326,465){\circle{12}}
\put(422,493){\circle{12}}
\put(518,519){\circle{12}}
\put(614,547){\circle{12}}
\put(710,573){\circle{12}}
\put(807,601){\circle{12}}
\put(903,626){\circle{12}}
\put(999,651){\circle{12}}
\put(1095,681){\circle{12}}
\put(1191,717){\circle{12}}
\put(1287,746){\circle{12}}
\put(1384,790){\circle{12}}
\put(230,435){\circle{12}}
\put(326,463){\circle{12}}
\put(422,490){\circle{12}}
\put(518,515){\circle{12}}
\put(614,541){\circle{12}}
\put(710,564){\circle{12}}
\put(807,589){\circle{12}}
\put(903,608){\circle{12}}
\put(999,625){\circle{12}}
\put(1095,644){\circle{12}}
\put(1191,665){\circle{12}}
\put(1287,671){\circle{12}}
\put(1384,683){\circle{12}}
\put(230,231){\circle*{12}}
\put(326,249){\circle*{12}}
\put(422,267){\circle*{12}}
\put(518,284){\circle*{12}}
\put(614,301){\circle*{12}}
\put(710,317){\circle*{12}}
\put(807,333){\circle*{12}}
\put(903,349){\circle*{12}}
\put(999,365){\circle*{12}}
\put(1095,383){\circle*{12}}
\put(1191,398){\circle*{12}}
\put(1287,416){\circle*{12}}
\put(1384,437){\circle*{12}}
\put(230,230){\circle{12}}
\put(326,248){\circle{12}}
\put(422,266){\circle{12}}
\put(518,282){\circle{12}}
\put(614,298){\circle{12}}
\put(710,313){\circle{12}}
\put(807,327){\circle{12}}
\put(903,342){\circle{12}}
\put(999,354){\circle{12}}
\put(1095,367){\circle{12}}
\put(1191,376){\circle{12}}
\put(1287,384){\circle{12}}
\put(1384,392){\circle{12}}
\put(230,232){\circle{12}}
\put(326,250){\circle{12}}
\put(422,268){\circle{12}}
\put(518,286){\circle{12}}
\put(614,303){\circle{12}}
\put(710,321){\circle{12}}
\put(807,338){\circle{12}}
\put(903,357){\circle{12}}
\put(999,376){\circle{12}}
\put(1095,398){\circle{12}}
\put(1191,421){\circle{12}}
\put(1287,448){\circle{12}}
\put(1384,483){\circle{12}}
\end{picture}
\caption{\it $Z_N/\mu^N$ versus $N$ at $\omega=1.0$ and
$\omega=1.1$. At each value of $\omega$, three different 
values of $\mu$ are shown. We estimate $\mu(1.0)=6.77404(2)$, and
$\mu(1.1)=6.89699(2)$. The filled circles correspond to the central
estimates of $\mu$ while open circles correspond to shifting $\mu$ by
the error estimate quoted.}
\label{figure5} 
\end{center}
\end{figure}

The swollen phase and the $\theta$-point behaviour can also be clearly
identified from the free-energy scaling. In the swollen phase we find
again the same behaviour as \cite{grassberger1994a-a}.  Figure
\ref{figure5} shows $Z_N/\mu^N$ for $\omega=1.0$ and $\omega=1.1$. For
each $\omega$, this quantity is plotted with three values of $\mu$
which differ in the sixth digit, showing both the accuracy in
the estimation of the free energy and the presence of logarithmic
corrections. At $\omega=1$, we estimate
$\mu(1)=\mu_{SAW}=6.77404(2)$. This can be compared to earlier
estimates of $6.7720(5)$ \cite{guttmann1978a-a}, $6.774(5)$
\cite{macdonald1992a-a}, and $6.77404(3)$ \cite{grassberger1994a-a}

\begin{figure}[p]
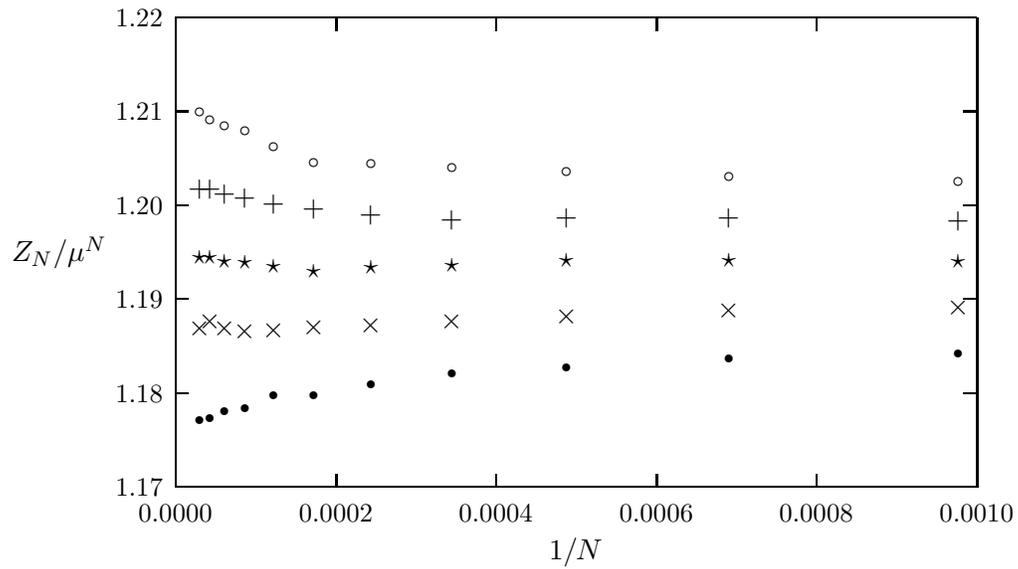

\begin{center}
\setlength{\unitlength}{0.240900pt}
\ifx\plotpoint\undefined\newsavebox{\plotpoint}\fi
\sbox{\plotpoint}{\rule[-0.200pt]{0.400pt}{0.400pt}}

\caption{\it $Z_N/\mu^N$ versus $1/N$ using our best estimates of
$\mu$ in the $\theta$-region:  
$\omega=1.180$, $1.181$, $1.182$, $1.183$, $1.184$ from top to
bottom.}
\label{figure6} 
\end{center}
\end{figure}

In the $\theta$-region, a similar analysis shows that here $Z_N$
scales as $\mu^N$ with weak $1/N$ corrections.  Figure \ref{figure6}
shows $Z_N/\mu^N$ plotted versus $1/N$, with respective values of
$\mu$ obtained in a similar fashion to the one shown in Figure
\ref{figure5}. We estimate the $\theta$-point to be
$\omega_\theta=1.182(1)$ and $\mu_\theta=7.011(2)$. (At fixed
$\omega$, the accuracy is of course higher: for $\omega=1.182$, we
estimate $\mu=7.0117675(5)$.)

\begin{figure}[p]
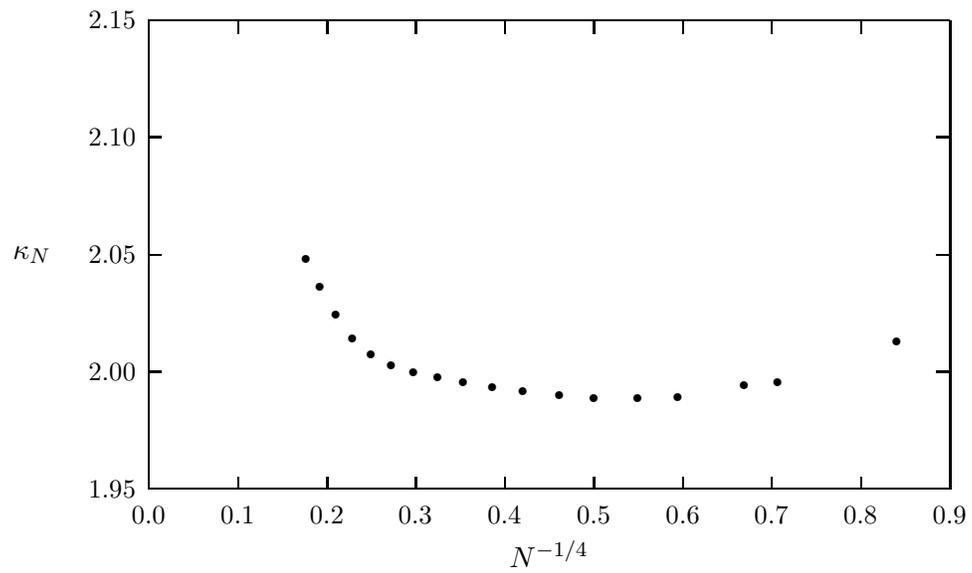

\begin{center}
\setlength{\unitlength}{0.240900pt}
\ifx\plotpoint\undefined\newsavebox{\plotpoint}\fi
\sbox{\plotpoint}{\rule[-0.200pt]{0.400pt}{0.400pt}}
%
\caption{\it Finite-size free energy $\kappa_N$ versus $N^{-1/4}$ for $\omega=1.4$.}
\label{figure7} 
\end{center}
\end{figure}

In the collapsed region, one expects the finite-size free energy to
have a strong correction term of the order $N^{-1/4}$ due to surface
effects. Figure \ref{figure7} shows this for $\omega=1.4$. As argued
above, the globule collapses when the length is above $N=250$, and we
notice here the onset of a corresponding strong change in the
behaviour of the finite-size free energy around this length
($N^{-1/4}\approx0.25$). Even though we cannot simulate long enough
chain lengths to clearly determine the precise nature of the
correction term, our data is certainly compatible with a $N^{-1/4}$
correction for $N^{-1/4}<0.2$.

\begin{figure}[p]
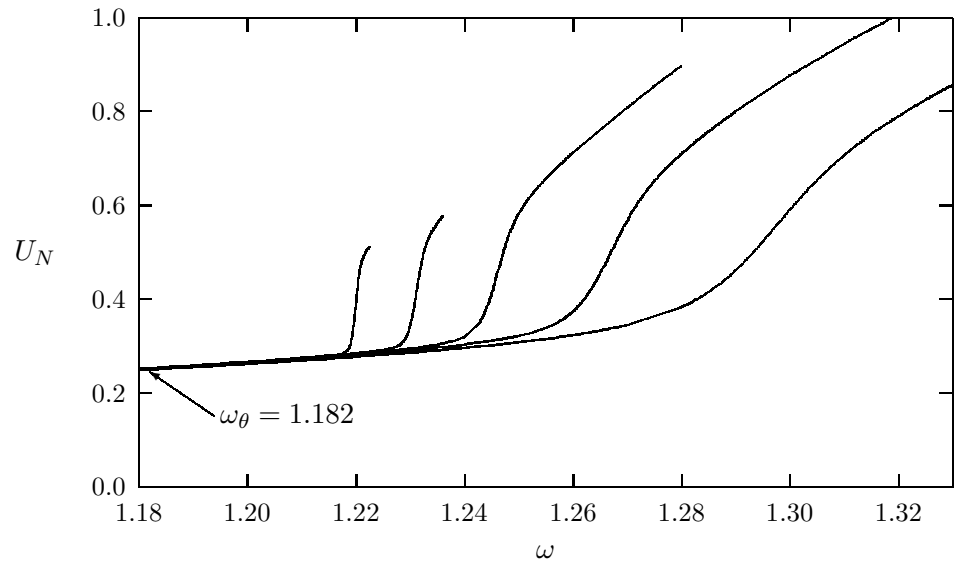

\begin{center}
\setlength{\unitlength}{0.240900pt}
\ifx\plotpoint\undefined\newsavebox{\plotpoint}\fi
\sbox{\plotpoint}{\rule[-0.200pt]{0.400pt}{0.400pt}}
%
\caption{\it Internal energy $U_N$ versus $\omega$ for lengths
$N=1024$, $2048$, $4096$, $8192$, and $16384$ from right to left
respectively, using the multi-histogram method. }
\label{figure8} 
\end{center}
\end{figure}

\begin{figure}[p]
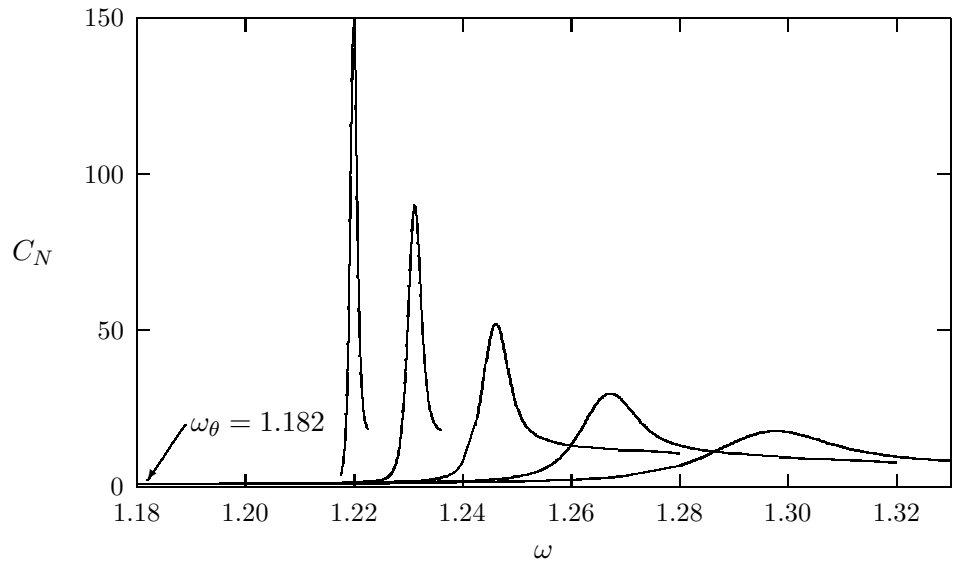

\begin{center}
\setlength{\unitlength}{0.240900pt}
\ifx\plotpoint\undefined\newsavebox{\plotpoint}\fi
\sbox{\plotpoint}{\rule[-0.200pt]{0.400pt}{0.400pt}}
%
\caption{\it Specific heat $C_N$ versus $\omega$ for lengths $N=1024$,
$2048$, $4096$, $8192$, and $16384$  from right to left
respectively, using the multi-histogram method.} 
\label{figure9} 
\end{center}
\end{figure}

In order to study the collapse transition more closely, we now focus
our attention on the internal energy and specific heat.  As can be
seen from Figures \ref{figure8} and \ref{figure9}, the internal energy
increases rapidly over a small temperature interval with a
corresponding diverging specific heat. As $N$ increases, the
transition becomes sharper and stays well separated from the
$\theta$-point, even though the location of the transition approaches
the $\theta$-point slowly.

\begin{figure}[p]
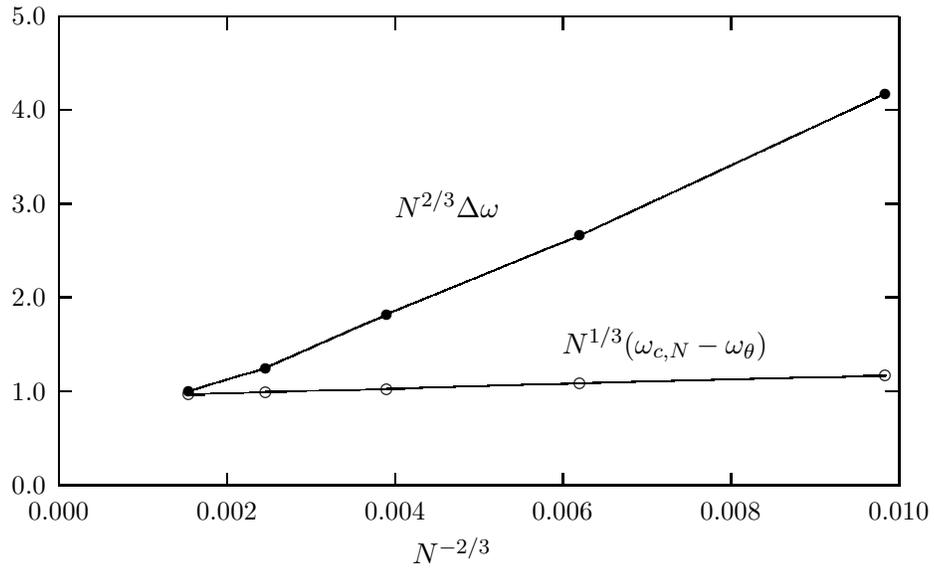

\begin{center}
\setlength{\unitlength}{0.240900pt}
\ifx\plotpoint\undefined\newsavebox{\plotpoint}\fi
\sbox{\plotpoint}{\rule[-0.200pt]{0.400pt}{0.400pt}}
%
\caption{\it Scaling of the transition: shift and width of the
collapse region. Shown are the scaling combinations 
$N^{1/3}(\omega_{c,N}-\omega_\theta)$ and $N^{2/3}\Delta\omega$ versus $N^{-2/3}$.} 
\label{figure10} 
\end{center}
\end{figure}

The scaling of the shift of the transition towards the $\theta$-point
$\omega_{c,N}-\omega_\theta$ and the sharpening of the transition
width $\Delta\omega$ are both shown in Figure \ref{figure10}. Here, we
defined the location of the collapse transition by the location of the
specific heat peak, and the width of the transition is given by the
interval in which the specific heat is greater or equal to half the
value of the peak height. Expecting from the KLG theory that
$\omega_{c,N}-\omega_\theta$ scales as $N^{-1/3}$ and that
$\Delta\omega$ scales as $N^{-2/3}$, we plot both
$N^{1/3}(\omega_{c,N}-\omega_\theta)$ and $\Delta\omega N^{2/3}$
versus $N^{-2/3}$ which was chosen empirically. Both quantities can be
seen to be asymptotic to constants: on the graph extrapolations give
non-zero intercepts. Hence, Figure
\ref{figure10} shows that the KLG predictions are compatible with our
simulations. We do note that the corrections to scaling for
$\Delta\omega$ are much larger than for
$\omega_{c,N}-\omega_\theta$.

\begin{figure}[p]
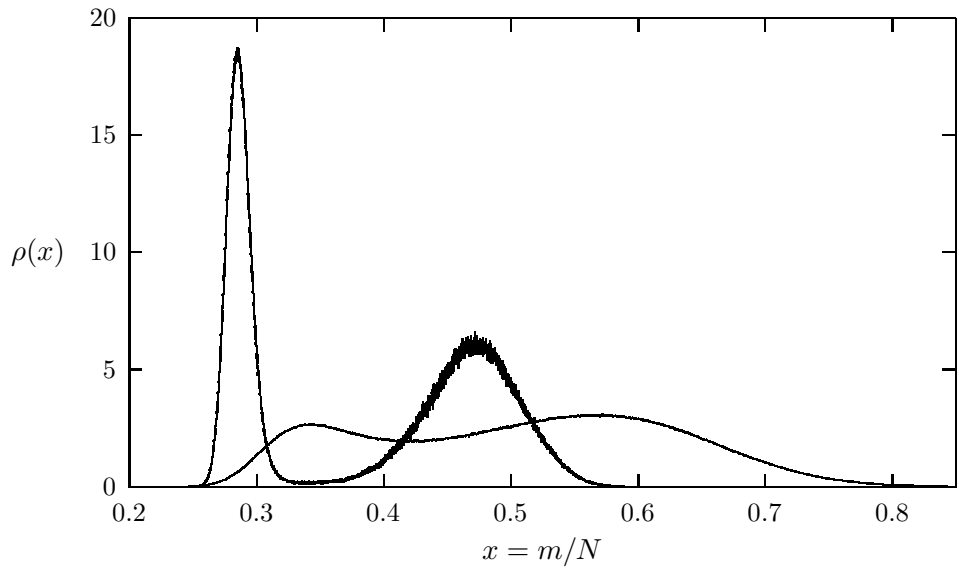

\begin{center}
\setlength{\unitlength}{0.240900pt}
\ifx\plotpoint\undefined\newsavebox{\plotpoint}\fi
\sbox{\plotpoint}{\rule[-0.200pt]{0.400pt}{0.400pt}}
%
\caption{\it Internal energy density distributions at $\omega_{c,N}$
for $2048$ and $16384$. The more highly peaked distribution
is associated with length $16384$.}
\label{figure11} 
\end{center}
\end{figure}

The character of the transition becomes apparent if one plots the
internal energy density distribution (rescaled density of
interactions) at the finite-size collapse transition,
$\omega_{c,N}$. Figure \ref{figure11} shows the emergence of a bimodal
distribution.  At length $2048$ one sees a slight non-convexity, which
at length $16384$ has evolved into a distribution dominated by two
sharp and well-separated peaks. The values of the minima and maxima of
the distribution are different by two orders of magnitude.

\begin{figure}[p]
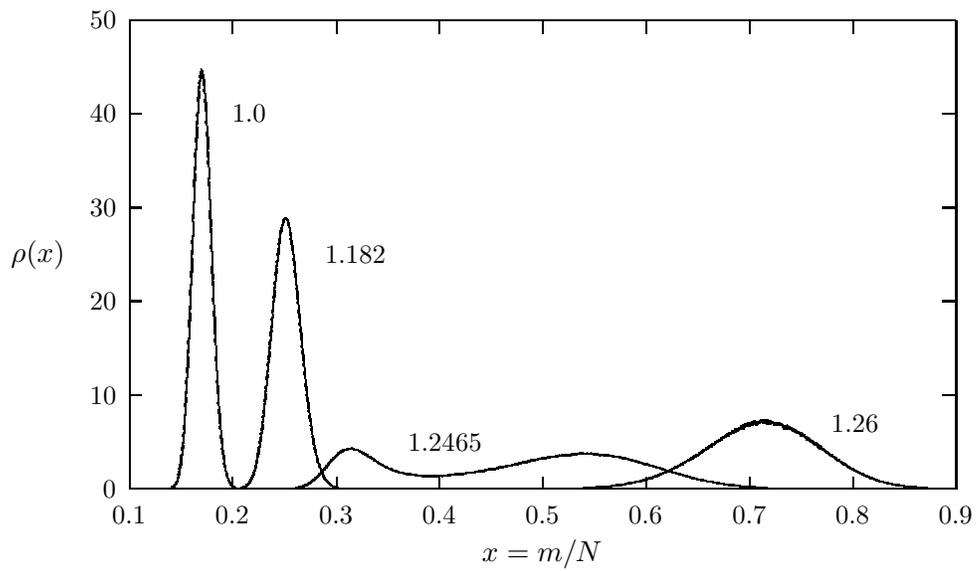

\begin{center}
\setlength{\unitlength}{0.240900pt}
\ifx\plotpoint\undefined\newsavebox{\plotpoint}\fi
\sbox{\plotpoint}{\rule[-0.200pt]{0.400pt}{0.400pt}}
%
\caption{\it Internal energy density distributions at $\omega=1.0$, $1.182$, $1.2465$ and $1.26$ for $N=4096$.}
\label{figure12} 
\end{center}
\end{figure}

It is instructive to study the transition by how this distribution
changes over a large range of $\omega$. Figure
\ref{figure12} shows this for $N=4096$. One sees that there is not
much of a change in the shape and location of the distribution between
the non-interacting case, $\omega=1$, and the $\theta$-point,
$\omega=1.182$. However, in a very small interval around the collapse
transition, the density distribution changes dramatically as $\omega$
increases. The density distribution switches from a peak located
around $0.3$ to a peak located around $0.55$, corresponding to a
sudden change in the internal energy. In the collapsed phase, the
width of the peak is much wider than in the swollen phase, implying a
larger specific heat. It is this difference between the swollen and
collapsed phases' specific heats that will eventually become the
thermodynamic second order jump.

\begin{figure}[p]
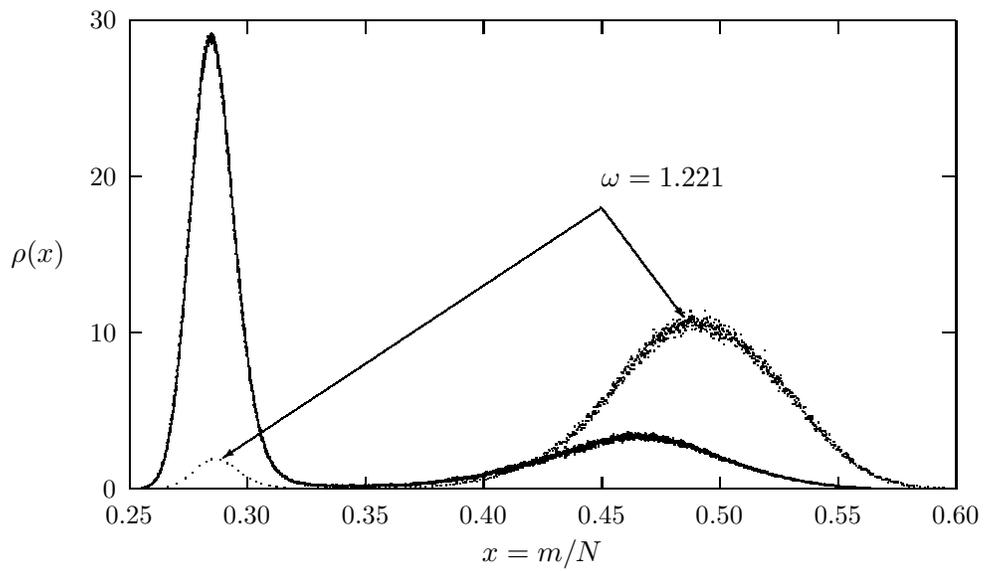

\begin{center}
\setlength{\unitlength}{0.240900pt}
\ifx\plotpoint\undefined\newsavebox{\plotpoint}\fi
\sbox{\plotpoint}{\rule[-0.200pt]{0.400pt}{0.400pt}}
%
\caption{\it Internal energy density distributions at $\omega=1.2195$ and $1.2210$ for $N=16384$.}
\label{figure13}
\end{center}
\end{figure}

The rapid first-order like switch between two peaks in the distribution
becomes more pronounced at larger polymer lengths.  At $N=16384$, this
`switching' is shown in Figure \ref{figure13}: over a range of
$\omega$ of the order of $10^{-3}$, a peak near $\omega=0.3$
disappears while another peak near $\omega=0.5$ emerges.

\begin{figure}[p]
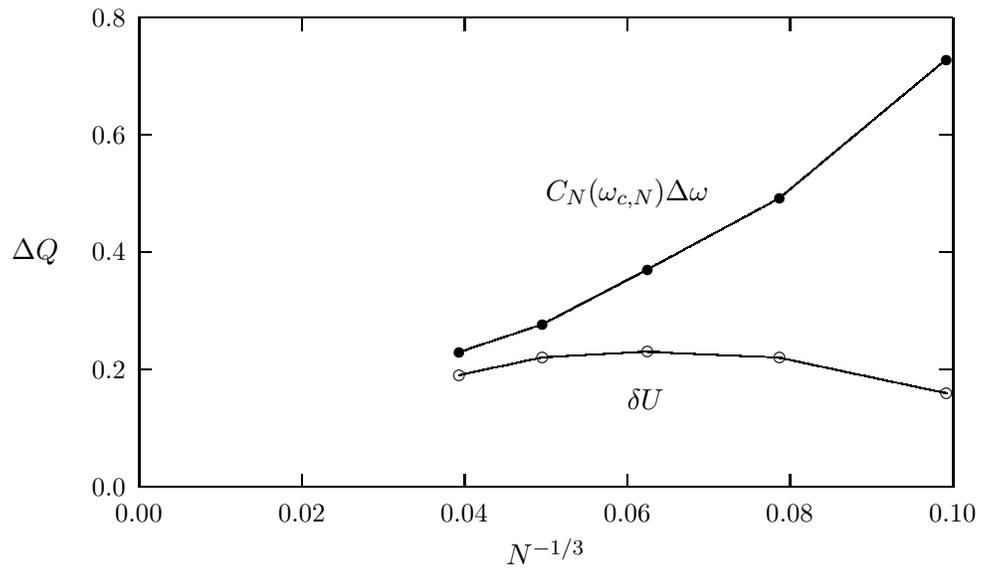

\begin{center}
\setlength{\unitlength}{0.240900pt}
\ifx\plotpoint\undefined\newsavebox{\plotpoint}\fi
\sbox{\plotpoint}{\rule[-0.200pt]{0.400pt}{0.400pt}}
%
\caption{\it Scaling of the latent heat $\Delta U$:
our two measures of $\Delta U$, $C_N(\omega_{c,N})\Delta\omega$ and
peak distance $\delta U$ are plotted versus $N^{-1/3}$.}
\label{figure14} 
\end{center}
\end{figure}

Returning to the scaling predictions from KLG theory, a suitably
defined finite-size latent heat, $\Delta Q$, should tend to zero as
$N^{-1/3}$ in the thermodynamic limit. One possible measure of this
latent heat is given by the product of specific heat peak
$C_N(\omega_{c,N})$ and specific heat width $\Delta\omega$, and
another is given by the distance $\delta U$ of the peaks in the
bimodal internal energy distribution.  Figure \ref{figure14} shows the
behaviour of both of these quantities.  One notices two things from
this figure. Firstly, we are unable to confirm or deny the predicted
scaling behaviour for $\Delta U$ and, secondly, even at length
$N=2048$ ($N^{-1/3}\approx 0.08$) there is considerable
discrepancy between the two quantities, so that it is not surprising
that one cannot discern a clear scaling behaviour.  The explanation
for the discrepancy between the two quantities as well as of the
difficulty of observing the predicted scaling behaviour is of course
that in order to observe the asymptotic behaviour the two peaks in the
histogram have to be well separated and distinct, and we see from
Figure \ref{figure11} that this is only the case when $N$ is of the
order of $10^4$. This explains why we are unable to find a value for
the exponent related to the divergence of the specific heat consistent
with the rest of our theoretical picture. We do concede that Figure
\ref{figure14} alone could be used to argue for the existence of a
real first-order transition in the thermodynamic limit, but we believe
the rest of our data and other theoretical facts provide a more
consistent picture.

In conclusion, our ISAW simulations elucidate the structure of the
polymer collapse transition in four dimensions.  We show conclusively
that there is indeed a collapse transition at a finite
temperature. Secondly, we find evidence for a $\theta$-temperature at
which the polymer is well approximated by Gaussian behaviour as well
as for a collapse transition which is well separated from the
$\theta$-point. The collapse transition shows many first-order like
features, such as a bimodal distribution in the internal energy. An
analysis of the scaling behaviour of this transition in the context of
the theory of Lifshitz, Grosberg and Khokhlov
\cite{lifshitz1978a-a,khokhlov1981a-a} shows that a consistent
interpretation of these findings is that of first-order like
finite-size corrections to a thermodynamic second-order transition. We
note that these findings are reminiscent of results for interacting
self-avoiding trails on the diamond lattice, where a $\theta$-point
was found in
\cite{prellberg1995b-:a} and subsequent simulations revealed a bimodal
distribution in the internal energy density  
\cite{grassberger1996a-a}. In \cite{prellberg1995b-:a} it was
concluded that the transition is second-order, whereas  
in \cite{grassberger1996a-a} the conclusion was that the transition is
first-order.  In the context of the findings presented here, it is
tempting to expect a similar resolution of this apparent contradiction
in terms of a pseudo-first-order transition.

\section*{Acknowledgements} 

Financial support from the Australian Research Council is gratefully
acknowledged by ALO while TP thanks the Department of Physics at the
University of Manchester and the Department of Mathematics and
Statistics at the University of Melbourne where parts of this work
were completed. This work was partially supported by EPSRC Grant No.\
GR/K79307, a visiting scholars award from the University of
Melbourne's Collaborative Research Grants scheme and the Australian
Research Council's small grants scheme. We thank A. J. Guttmann for
many useful comments on the manuscript.

\newpage 

\clearpage

\clearpage

\clearpage

\clearpage

\clearpage

\clearpage

\clearpage

\clearpage

\clearpage

\clearpage

\clearpage

\clearpage

\clearpage

\clearpage

\end{document}